\documentclass{article}
\usepackage{arxiv}
\def\bSig\mathbf{\Sigma}

\usepackage{amsmath}
\usepackage{amsfonts}
\usepackage{graphicx}
\usepackage{subcaption}
\usepackage{booktabs, caption, longtable, colortbl, array}
\usepackage{url}
\usepackage{natbib}
\usepackage{bm}
\usepackage{float}
\usepackage{soul}

\title{Supervised Integrative Biclustering with applications to Alzheimer's Disease}

\author{{\hspace{1mm}Kaifeng Yang} \\
	Division of Biostatistics and Health Data Science\\
	University of Minnesota Twin Cities\\
	Minneapolis, MN 55455 \\
	\texttt{yang7521@umn.edu} \\
	\And
	{\hspace{1mm}Thierry Chekouo} \\
	Division of Biostatistics and Health Data Science\\
	University of Minnesota Twin Cities\\
	Minneapolis, MN 55455 \\
	\texttt{tchekouo@umn.edu} \\
        \And
	{\hspace{1mm}Sandra E. Safo}\thanks{Corresponding Author: Sandra E. Safo, www.sandraesafo.com} \\
	Division of Biostatistics and Health Data Science\\
	University of Minnesota Twin Cities\\
	Minneapolis, MN 55455 \\
	\texttt{ssafo@umn.edu} \\
}

\begin{document}
\maketitle

\label{firstpage}


\begin{abstract}
Multiple types or views of data (e.g. genetics, proteomics) measured on the same set of individuals are now popularly generated in many biomedical studies. A particular interest might be the detection of sample subgroups (e.g. subtypes of disease) characterized by specific groups of variables. Biclustering methods are well-suited for this problem since they can group samples and variables simultaneously. However, most existing biclustering methods cannot guarantee that the detected sample clusters are clinically meaningful and related to a clinical outcome because they independently identify biclusters and associate sample clusters with a clinical outcome. Additionally, these methods have been developed for continuous data when integrating data from different views and do not allow for a mixture of data distributions. We propose a new formulation of biclustering and prediction method for multi-view data from different distributions that enhances our ability to identify clinically meaningful biclusters by incorporating a clinical outcome. Sample clusters are defined based on an adaptively chosen subset of variables and their association with a clinical outcome. We use extensive simulations to showcase the effectiveness of our proposed method in comparison to existing methods. Real-world applications using lipidomics, imaging, and cognitive data on Alzheimer's disease(AD) identified biclusters with significant cognitive differences that other methods missed. The distinct lipid categories and brain regions characterizing the biclusters suggest potential new insights into pathology
of AD.

\end{abstract}

%
\keywords{Bicluster \and Multi-view Data Integration \and Interpretable \and Subgroup Detection \and Supervised.}



%

\section{Introduction}
\label{s:intro}

\par Alzheimer's disease (AD) is a complex and heterogeneous disease that affects more than 7 million adults in the U.S. \citep{alzheimersassociation2024AlzheimersDisease2024}. Several factors such as age, genetics, and behavior affect the risk and progression of AD, but the effects of these factors differ between individuals \citep{silvaAlzheimersDiseaseRisk2019, avelar-pereiraDecodingHeterogeneityAlzheimers2023}. The heterogeneity of AD has spurred research aimed at identifying subgroups of AD with different risk profiles based on phenotypic and/or multiomics data \citep{wangSupervisedConvexClustering2023,zhangRobustKnowledgeguidedBiclustering2024,elmanExploringGeneticHeterogeneity2024}. Finding subgroups in the
population that have a similar pathology, interpreting these subgroups by clinical outcomes,
and identifying risk factors that characterize these subgroups can provide a deeper understanding of the
risk of AD between different subgroups.

The Alzheimer's Disease Neuroimaging Initiative (ADNI) is a popular study that supports the investigation and development of treatments that slow or stop AD progression by providing valuable data from older participants with or without AD. Data from multiple sources, including genetic, omics, imaging, clinical, demographic, and outcome measures (such as AD status, Mini Mental State Examination (MMSE), and Clinical Dementia Rating – Sum of Boxes (CDR-SB) for AD progression), were collected. These datasets can be used to identify subgroups within AD and deepen our understanding of its heterogeneity. Recent studies have shown promising results using molecular and clinical data from ADNI to identify AD subgroups characterized by molecular data for a better understanding of the heterogeneity of AD \citep{ elmanExploringGeneticHeterogeneity2024, zhangRobustKnowledgeguidedBiclustering2024}

\par Statistically, simultaneously identifying subgroups and variables that characterize the subgroups is a biclustering problem. Biclustering, first introduced in 1972 \citep{hartiganDirectClusteringData1972}, has become a popular unsupervised statistical method for identifying subgroups of diseases, especially for high-dimensional biomedical data. Biclustering simultaneously identifies sets of samples characterized by sets of variables, and each set of samples and variables forms a bicluster. Unlike clustering methods, which group all samples based on the entire set of variables, biclustering focuses on the relevant variables for each bicluster, enabling the discovery of patterns that might remain hidden when using all variables. Biclustering methods have been developed for subgroup detection based on single data type [e.g., Sparse Singular Value Decomposition incorporating Stablility Selection (S4VD) \citep{sillRobustBiclusteringSparse2011}, the penalized Plaid \citep{chekouoPenalizedBiclusteringModel2015} and Gibbs-Plaid \citep{chekouoGibbsplaidBiclusteringModel2015}] or multiple data types [e.g., Bayesian generalized biclustering analysis via adaptive structured shrinkage (GBC) \citep{ liBayesianGeneralizedBiclustering2018} and robust integrative biclustering for multi-view data (iSSVD) \citep{zhangRobustIntegrativeBiclustering2022}]. These methods have been applied to a wide variety of complex diseases and have shown the potential to provide biological and clinical insights based on molecular and clinical data \citep{castanhoBiclusteringDataAnalysis2024,elmanExploringGeneticHeterogeneity2024}.

\par Most existing biclustering methods do not incorporate an outcome when generating biclusters \citep{wangSupervisedConvexClustering2023}. A typical approach involves a two-step process: first, identifying the biclusters, and then linking them to an outcome (e.g., \cite{elmanExploringGeneticHeterogeneity2024, zhangRobustIntegrativeBiclustering2022,yangAnalysisBreastCancer2017, sillRobustBiclusteringSparse2011}). However, this approach can be problematic because the steps are independent and the identified biclusters may not be associated with any outcome, making them potentially clinically meaningless. As an example, we applied iSSVD \citep{zhangRobustIntegrativeBiclustering2022} and S4VD \citep{sillRobustBiclusteringSparse2011} to molecular data from the ADNI study to detect biclusters. Unfortunately, the identified biclusters were not associated with AD risk or progression, rendering them clinically meaningless. This highlights a limitation of traditional biclustering methods, where the results may fail to provide clinically relevant insights unless an outcome is directly integrated into the biclustering step.

\par To our knowledge, Supervised Convex Clustering (SCC) \citep{wangSupervisedConvexClustering2023} is the only biclustering method that takes advantage of an outcome to aid in the identification of biclusters. However, SCC and many other biclustering methods are designed to analyze data from a single source and are not suitable for multi-source or multi-view data, which are now commonly available in many biomedical studies. Furthermore, these methods assume that the data follow a normal distribution, limiting their flexibility to handle various data types (e.g. binary, count). Notably, simply stacking multiple types of data and applying biclustering methods developed for a single data source and type does not rigorously model the dependencies among views. Additionally, this approach may lead to the detection of biclusters characterized by variable clusters that are based on only a few views, missing richer, integrated information across all views \citep{sunMultiviewSingularValue2014, zhangRobustIntegrativeBiclustering2022}. Effective analysis of multi-view data requires considering both individual views and their connections, enabling more complex and insightful models that reveal underlying structures more effectively.

\par In this paper, we propose a novel method, Supervised Integrative Biclustering (SIB) for multi-view data that borrows strengths from an outcome when integrating data from multiple sources to detect biclusters. We make three main contributions to integrative biclustering. First, we incorporate outcome supervision to guide bicluster identification--a feature rarely explored in existing biclustering methods for multi-view data--enhancing the interpretation of the identified biclusters.
Second, our formulation is flexible, allowing multiple types of data with different distributions (e.g., Gaussian for continuous data, Bernoulli for binary data, etc.) to be used to identify biclusters. Third, we develop a computationally efficient algorithm in Python and interface with the R programming language, for ease of use.  Our algorithm is faster than existing methods that account for various types of data distributions such as GBC \citep{liBayesianGeneralizedBiclustering2018}. The rest of the paper is organized into the following sections: In Section 2, we describe our proposed method; in Section 3, we describe our simulation settings and compare our methods with other existing methods; in Section 4, we apply our proposed methods to find subgroups of Alzheimer's disease (AD); in Section 5, we briefly summarize our method and discuss its limitations and possible extensions. 

\section{Model}
\label{s:model}

\subsection{General Notations}
\par Assume that $d=1,\dots, D$ views of data are available. Denote the data in the $d^{\text{th}}$ view as $\bm{X}^{(d)}\in \mathbb{R}^{n\times p^{(d)}}$, where $n$ is the sample size of the data and $p^{(d)}$ is the number of variables in the $d^{\text{th}}$ view, and let $\bm{X}=\{\bm{X}^{(1)},\dots,\bm{X}^{(D)}\}$ be the collection of the data from all views. Let $\bm{y}$ be a vector of the outcome, where each entry $y_i$ is the outcome of the $i^\text{th}$ individual, $i=1,...,n$. In our motivating study, $\bm{X}^{(1)}$ and $\bm{X}^{(2)}$ consist of lipidomics and imaging data, respectively, and $\bm{y}$ is a cognitive score, the Mini Mental State Examination (MMSE) \citep{kurlowiczMiniMentalStateExamination1999}. In addition, we use $\bm{X_E}$ to denote covariates that could be directly associated with the outcome (e.g., age) but may not be of direct interest as variable clusters. We assume each entry of the data in each view and the outcome come from a single-parameter exponential family distribution to allow more flexible options for different data types. However, to illustrate our method, we assume that the entries in each view are continuous while the outcome is assumed to be either continuous or binary. The probability density function of each entry $a$ from $\bm{X}^{(d)}$ or $\bm{y}$ can be expressed as $f(a|\psi)=p_0(a)\text{exp}\{a\psi - G(\psi)\}$ and $\log f(a|\psi)=\log p_0(a)+a\psi - G(\psi)$, where $\psi$ is the natural parameter of the distribution, $G(a)$ is a convex cumulant function, $p_0(a)$ is a non-negative normalization function, and the expectation $E(a|\psi)=G'(\psi)$. The choice of function $G$ for common distributions from the exponential family is summarized in Web Table A.1 in the supporting information. Let $\bm{\Psi}^{(d)}\in \mathbb{R}^{n\times p^{(d)}}$ be a matrix that collects all the natural parameters in the $d^{\text{th}}$ view and $\bm{\psi_y}\in \mathbb{R}^{n}$ be a vector of all the natural parameters for $\bm{y}$. We further assume that the entries in $\bm{X}^{(d)}$ and $\bm{y}$ are conditionally independent given the natural parameters, so the likelihood of $\bm{X}$ and $\bm{y}$ is the multiplication of the likelihood of each individual component.

\par Without further specification, we denote the $i^{\text{th}}$ row and $j^{\text{th}}$ column of a matrix $\bm{A}$ as $\bm{A}_{i\cdot}$ and $\bm{A}_{\cdot j}$, respectively, and the entry on the $i^{\text{th}}$ row and $j^{\text{th}}$ column as $a_{ij}$ throughout this paper. The $i^{\text{th}}$ entry of a vector $\bm{a}$ is denoted as $a_i$, and the $p$-norm of a vector is defined as $|\bm{a}|_p=\sqrt[\leftroot{-2}\uproot{2}p]{\sum_{i=1}^n |a_i|^p}$. For example, $|\bm{a}|_2=\sqrt{\sum_{i=1}^n |a_i|^2}$, and $|\bm{a}|_1=\sum_{i=1}^n |a_i|$. $\bm{A}^T$ denotes the transpose of a matrix or a vector $\bm{A}$.

\subsection{Biclustering for multiple views}
\label{biclustering_data}

\par  We model the data $\bm{X}^{(1)},\dots, \bm{X}^{(D)}$ from all views through the natural parameter $\bm{\Psi}^{(d)}$, which is linked with the latent components as $\bm{\Psi}^{(d)} = \bm{1\mu}^{(d)T}+(\bm{U}\cdot \bm{W})\bm{V}^{(d)T}$, where $\cdot$ represents elementwise matrix multiplication. In this formulation, we assume that there are $K$ biclusters and integrate the data $\bm{X}^{(d)}$ from all views using a low-dimensional common score $\bm{U}\in \mathbb{R}^{n\times K}$ that acts as a shared driver among all views and data-specific loadings $\bm{V}^{(1)}, \dots, \bm{V}^{(D)}$ for each view, where each $\bm{V}^{(d)}\in \mathbb{R}^{p^{(d)}\times K}$. The matrix $\bm{W}$ is a sample bicluster indicator matrix with $w_{ik}$ being 1 if the $i^{\text{th}}$ sample belongs to the $k^{\text{th}}$ bicluster, and 0 otherwise. $\bm{\mu}^{(d)}$ is a view-specific intercept that accounts for the difference in mean of the variables. The variable bicluster for each view is given by data-specific loadings $\bm{V}^{(d)}$. If variable overlapping is allowed among biclusters, then each non-zero term $v_{jk}^{(d)}$ in $\bm{V}^{(d)}$ indicates that the $j^\text{th}$ variable in view $d$ belongs to the $k^\text{th}$ bicluster. Otherwise, if we assume that each variable can only belong to at most one bicluster, then the entry with the maximum nonzero absolute value in each row indicates the bicluster that this variable belongs to. If all the entries in a row are zero, then the variable does not belong to any bicluster and is deemed unimportant. 
\par We construct the loss function for biclustering data from multiple views using the negative log-likelihood. Since the entries of the data in each view are assumed to come from an exponential family distribution and are conditionally independent, the loss function in the data matrices $\bm{X}^{(d)}$ from all views can be expressed in the following form: $\log(P(\bm{X}^{(d)}|\bm{\Psi}^{(d)}))=\sum_{i=1}^n\sum_{j=1}^{p^{(d)}}\log P(x_{ij}^{(d)}|\psi_{ij}^{(d)})=\sum_{i=1}^n\sum_{j=1}^{p^{(d)}}(-\log P_0(x_{ij}^{(d)})-x_{ij}^{(d)}\psi_{ij}^{(d)}+G(\psi_{ij}^{(d)}))$, where $G(\psi_{ij}^{(d)})$ is a convex function corresponding to the distribution specific to view $\bm{X}^{(d)}$. In addition, we add an $L_1$ penalty $ \sum_{k=1}^{K} \lambda_{kd}\sum_{j=1}^{p^{(d)}}|v^{(d)}_{jk}|$ on each column of $\bm{V}^{(d)}$ to encourage sparsity in each column and select important variables within each bicluster. In practice, we also provide an option to use the same $\lambda_{kd}$ for all components within the same view to reduce the number of hyper-parameters to tune, and the penalty term becomes $\lambda_{d}\sum_{k=1}^{K} \sum_{j=1}^{p^{(d)}}|v^{(d)}_{jk}|$. 
\par In this formulation, we also impose the following constraints on $\bm{U}$, and $\bm{W}$: (1) the columns of $\bm{U}$ satisfy $|\bm{U}_{\cdot k}|_2=1$; (2) the rows of $\bm{W}$ satisfy $w_{ik}\in\{0,1\}$ and $\sum_{k=1}^K w_{ik}=1$ for all $i\in \{1,\dots,n\}$. The first constraint helps to ensure stability and smoothness of estimation and prevent computational problem caused by extremely huge $\bm{\Psi}^{(d)}$. The second constraint ensures that each sample belongs to exactly one bicluster. Although this constraint can be implemented using algorithms such as mixed integer programming, such algorithms tend to be computationally inefficient, so we relax it to an alternative version: each row $i$ of $\bm{W}$ satisfy $ 0\leq w_{ik}\leq 1$ for all $k\in \{1,\dots, K\}$ and $\sum_{k=1}^K w_{ik}=1$, as done in fuzzy clustering and soft assignments \citep{miyamotoBasicMethodsCMeansClustering2008,wangLaplacianKmodesAlgorithm2014}. This is easier and faster to implement and provides an alternative interpretation to $\bm{W}$. Each entry $w_{ik}$ is the probability that the $i^\text{th}$ individual belongs to the $k^\text{th}$ bicluster. We assign each sample $i$ to the bicluster $\widehat{k}=\text{argmax}_kw_{ik}$ so that each sample belongs to exactly one bicluster.

\subsection{Supervised Biclustering for multiple views}
\par One important aspect of our method that sets us apart from most existing methods is the inclusion of an outcome in determining the biclusters. This provides a natural interpretation of the biclustering results without further steps, thus yielding clinically meaningful biclusters. These outcomes can be of various types and serve to interpret the identified biclusters. For instance, we used the MMSE, a continuous cognitive test score, to guide the detection of biclusters among individuals at risk or having AD \citep{kurlowiczMiniMentalStateExamination1999}. Furthermore, in our proposed method, the outcome $\bm{y}$ is modeled using the natural parameter $\bm{\psi_y}$ from exponential families, and we model $\bm{\psi_y}=\bm{W\beta}+\bm{X}_E\bm{\beta}_E$, where $\bm{W}$ is the bicluster indicator matrix defined in Section \ref{biclustering_data} and $\bm{X}_E$ are optional additional covariates associated with the outcome. Note that we can concatenate $\bm{W}$ and $\bm{X}_E$ to estimate the concatenation of $\bm{\beta}$ and $\bm{\beta}_E$, so we describe our estimation and prediction methods without the presence of these additional covariates $\bm{X}_E$ for simplicity. The estimated $\bm{\beta}$ can be used to interpret and compare the biclusters with respect to the outcome. We can also use the estimated $\bm{\beta}$ to predict the outcome $y$, which can be interpreted as the expectation of the outcome for an individual. Additionally, $\bm{\beta}$ can be used to align biclustering results from different datasets. For instance, after obtaining a biclustering result on a training set, it is common to use a separate testing dataset to validate the result. One challenge in comparing the testing and training results is the alignment of the bicluster labels since the first bicluster in the testing results may not correspond to the first one in the training results. In this case, $\bm{\beta}$ can provide strong evidence of whether or not two biclusters from different datasets correspond to each other by comparing the value of $\bm{\beta}$ of each bicluster.

\par Similarly, since the outcome $\bm{y}$ is from an exponential family distribution, we can construct a loss function for $\bm{y}$ using the negative log-likelihood by $\log P(\bm{y}|\bm{\psi_y}) = \sum_{i=1}^n(-\log P_0(y_i)+y_i{\psi_y}_i-G({\psi_y}_i))$, where $G({\psi_y}_i)$ is a convex function corresponding to the data type of $\bm{y}$.

\par Combining the loss function of $\bm{y}$, and the loss function of $\bm{X}$, we propose a total loss function for all views and the outcome as:
\begin{align*}
    l(\bm{X,y,U,V,W,\mu,\beta})=&(1-\rho)\frac{1}{n}\sum_{i=1}^n(y_i{\psi_y}_i-G({\psi_y}_i))\\
    &+ \rho\sum_{d=1}^D\sum_{i=1}^n\sum_{j=1}^{p^{(d)}}\frac{1}{n\times p^{(d)}}(-x_{ij}^{(d)}\psi_{ij}^{(d)}+G(\psi_{ij}^{(d)}))\\
    &+ \sum_{d=1}^{D}\sum_{k=1}^{K} \lambda_{kd}\sum_{j=1}^{p^{(d)}}|V^{(d)}_{jk}|.
\end{align*} 

\par The terms $\sum_{i=1}^n\sum_{j=1}^{p^{(d)}}\log P_0(x_{ij}^{(d)})$ and $\sum_{i=1}^n\log P_0(y_i)$ in the loss function are omitted for simplicity because they are constants with respect to the parameters we estimate. To balance the magnitude of the supervision and biclustering terms, we divide each part by the size of the summation (i.e. divide the biclustering term by $n\times p^{(d)}$ and the supervision term by $n$) and use $\rho$ to control the weights of the loss for these terms to account for the fact that $\bm{X}$ and $\bm{y}$ may not be on the same scale. 

\subsection{Parameter Estimation}
\label{ss: hyper}
\par The parameters are estimated using alternating projected gradient descent \citep{bubeckConvexOptimizationAlgorithms2015} such that $(\widehat{\bm{U}},\widehat{\bm{V}},\widehat{\bm{W}},\widehat{\bm{\mu}},\widehat{\bm{\beta}})=\text{argmin }l(\bm{X,y,U,V,W,\mu,\beta})$. An algorithm to estimate the parameters is included in the supporting information Section D. Following ideas in \cite{liBayesianGeneralizedBiclustering2018}, we initialize each entry in $\bm{\Psi}^{(d)}$ as $\Psi_{ij}^{(d)}=h(x_{ij}^{(d)})$, where the choice of the function $h(a)$ depends on the distribution of the data type in the  $d^{\text{th}}$ view and is summarized in the supporting information Section A.2.

Denote the estimate of each parameter $\bm{A}$ at $m^\text{th}$ iteration as $\bm{A}_{(m)}$, and let the initialization be $\bm{A}_{(0)}$. We initialize $\bm{U}_{(0)}$ and $\bm{V}^{(d)}_{(0)}$ as follows: Let $\bm{\Psi} = [\bm{\Psi}^{(1)},\dots,\bm{\Psi}^{(D)}]$ be the concatenation of $\bm{\Psi}^{(d)}$ in all views and $\bm{\Psi}=\bm{P\Sigma Q}^T$ be the singular value decomposition of $\bm{\Psi}^{(d)}$, where $\bm{P}$ and $\bm{Q}$ are the left and right singular matrices, and $\bm{\Sigma}$ is the diagonal matrix of singular values. Set $\bm{U}_{(0)}$ as the first $K$ columns of $\bm{P}$, $\bm{V}_{(0)}$ as the first $K$ columns of $\bm{V}_{(0)}=\bm{Q\Sigma}$, and divide $\bm{V}_{(0)}$ into $\bm{V}^{(d)}_{(0)}$ based on the number of variables in each view. $\bm{W}_{(0)}$ is initialized as a matrix of ones, and $\bm{\mu}_{(0)}$ and $\bm{\beta}_{(0)}$ are initialized as matrices of zeros.

\par Each parameter is updated using the projected gradient descent algorithm with a pre-specified step size $\alpha$ that controls the convergence rate. Denote the differentiable part of the loss function as $l_0(\bm{X,y,U,V,W,\mu,\beta})=\rho\sum_{d=1}^D\sum_{i=1}^n\sum_{j=1}^{p^{(d)}}\frac{1}{n\times p^{(d)}}(-x_{ij}^{(d)}\psi_{ij}^{(d)}+G(\psi_{ij}^{(d)})) + (1-\rho)\frac{1}{n}\sum_{i=1}^n(y_i{\psi_y}_i-G({\psi_y}_i))$. The penalty term in the loss function is accounted for by the proximal map when updating $\bm{V}$. For each parameter $\bm{A}$, we update them by $\Tilde{\bm{A}}_{(m+1)}=(\bm{A}_{(m)}-\alpha  \nabla l_0/\nabla \bm{A}_{(m)})$, where $\nabla l_0/\nabla \bm{A}_{(m)}$ denotes the gradient of $l_0$ with respect to the parameter $\bm{A}$. We use $\Tilde{\bm{A}}$ to denote the unprojected parameter estimate, and $\bm{A}\in\{\bm{U,V,W,\mu,\beta}\}$. If the updated parameter has a constraint associated with it, then we will project the parameter using their corresponding projections as follows: Each column of $\bm{U}$ is projected using $\bm{U}_{\cdot,k,(m+1)}=\Tilde{\bm{U}}_{\cdot,k,(m+1)}/|\Tilde{\bm{U}}_{\cdot,k,(m+1)}|_2$ to ensure that $|\bm{U}_{\cdot,k,(m+1)}|_2=1$. We project each entry of $\bm{V}$ by the proximal map of Lasso penalty $\bm{V}^{(d)}_{j,k,(m+1)}=S(\Tilde{\bm{V}}^{(d)}_{j,k,(m+1)},\lambda_{kd}\alpha)$, where $S(x, \lambda)$ is the soft-thresholding function described in supporting information Section A.2. Each row of $\bm{W}$ is projected onto a probability simplex using the algorithm proposed by \cite{wangProjectionProbabilitySimplex2013}. After all the parameters are converged, since the estimation of $\bm{\beta}$ depends only on the estimation of $\bm{W}$, we use the objective ${\arg\min}_{\bm{\beta}} \sum_{i=1}^n(y_i{\psi_y}_i-G({\psi_y}_i))$ to estimate $\bm{\beta}$, for better results.

\par The hyper-parameters $\lambda_{kd}$'s are selected using the Bayesian information criterion (BIC) or the extended BIC (eBIC) proposed by \cite{chenExtendedBayesianInformation2008}. The number of biclusters $K$ is selected as the maximum $K$ that does not produce an empty bicluster if the user does not specify it. We provide details on the choice of BIC and eBIC as well as steps to choose $K$ in the supporting information Section A.3.

\subsection{Prediction}
\label{ss: pred}
\par Another important aspect of our method is that it can be used to predict bicluster assignment and outcomes for new samples from the same population. If new samples $\bm{X}_\text{test}=\{\bm{X}_\text{test}^{(1)},\dots,\bm{X}^{(D)}_\text{test}\}$ are available, we can use the estimated $\bm{\widehat{V}}$ and $\bm{\widehat{\mu}}$ to estimate $\bm{W}$ for the new samples since $\bm{\widehat{V}}$ and $\bm{\widehat{\mu}}$ are variable-level estimators and should be the same for any new samples from the same population. Also, since the estimation of $\bm{W}$ depends on $\bm{U}$, we need to estimate both. We assume that the new samples do not have the outcome information, so we only use the loss function for multiple views of data, defined as $l_x(\bm{X,U,V,W,\mu})=\sum_{i=1}^n\sum_{j=1}^{p^{(d)}}\frac{1}{n\times p^{(d)}}(-x_{ij}^{(d)}\psi_{ij}^{(d)}+G(\psi_{ij}^{(d)}))$, to estimate $\bm{W}$ and $\bm{U}$. They are estimated such that $\bm{\widehat{U}', \widehat{W}'}=\text{argmin }l_x(\bm{X_\text{test},U,\widehat{V},W,\widehat{\mu}})$ under the same constraints for $\bm{W}$ and $\bm{U}$. Then, we can use the estimated $\bm{\widehat{W}'}$ from new samples and the estimated $\bm{\widehat{\beta}}$ from the bicluster results to estimate the natural parameter $\bm{\psi}_y'=\bm{\widehat{W}'}\bm{\widehat{\beta}}$ and use the corresponding distribution to predict the outcomes.

\section{Simulation}
\label{s:sim}

\subsection{Simulation Design}
\par To evaluate the performance of our methods, we design a series of simulations with different dimensions and distributions of the outcomes and compare our performance with other existing methods. Three sets of sample sizes and number of variables, namely $(n=150, p^{(d)}=100)$, $(n=150, p^{(d)}=500)$, and $(n=500, p^{(d)}=1000)$, are generated for outcomes from Gaussian and Bernoulli distributions, respectively, under the assumption in section 2.2 and 2.3. For the multi-view data $\bm{X}$, we assumed data from all views to be continuous and the entries follow a Gaussian distribution in all cases. In addition, we generate an independent testing set of the same size as the training data to assess the performance of predicting the outcome on new samples. We choose hyper-parameters, including the number of biclusters, using methods described in the supporting information Section A.3. We describe our simulation settings in detail in the supporting inforamtion Section B.1.

\par We compare our methods with Generalized Biclustering (GBC) \citep{liBayesianGeneralizedBiclustering2018}, sparse SVD algorithm with nested stability selection (S4VD) \citep{sillRobustBiclusteringSparse2011}, and integrative sparse singular value decomposition (iSSVD) \citep{zhangRobustIntegrativeBiclustering2022}. We include a brief summary of these methods in the supporting information Section B.2. The tuning parameters were selected using the method described in each of these methods. None of the comparison methods incorporates the outcome when identifying biclusters. 

\par For each of the comparison methods, we create a matrix $\bm{W}$ that represents the sample assignment from the bicluster identified such that $w_{ik}=1$ if sample $i$ is in the $k^{\text{th}}$ bicluster and $w_{ik}=0$ otherwise. This matrix is used to fit a linear regression model for the continuous outcome and a logistic regression model for the binary outcome. We use the regression models to evaluate how well the comparison methods fit the outcomes. We use the nearest centroid algorithm to assign biclusters to the testing samples since none of them provide a prediction method and use the estimate from the regression models to predict the test outcome $\bm{y}$.

\subsection{Evaluation Methods}
\label{ss:evaluation}
\par We consider multiple evaluation metrics to assess the performance of our models and the comparison methods comprehensively. These evaluation metrics can be categorized as measuring the accuracy of biclusters and the outcome. We use relevance, recovery, F-score, false positive, and false negative to evaluate the accuracy of the biclusters obtained. These metrics are calculated using the Jaccard similarity matrix \citep{sillRobustBiclusteringSparse2011, zhangRobustIntegrativeBiclustering2022, sunNoiseResistantBiclusterRecognition2013}. Relevance is calculated as the average of the maximum Jaccard similarity of each true bicluster across all the identified biclusters, and recovery is the average of the maximum Jaccard similarity of each identified bicluster across all the true biclusters. Relevance and recovery measure how well the true bicluster is represented and recovered by the identified bicluster, respectively. The F-score summarizes both metrics by the harmonic mean of relevance and recovery. False positive and false negative measure the proportion of elements selected by the identified bicluster that are not in the true bicluster and the proportion of elements not selected by the identified bicluster that are in the true bicluster, respectively. We describe these metrics in greater detail in the supporting information Section B.3. The accuracy of the outcomes is assessed differently depending on the type of outcome. We use mean square error (MSE) and error rate to evaluate the accuracy of continuous and binary outcomes, respectively, for both training and testing data. Since comparison methods do not predict the outcome $\bm{y}$, we use linear or logistic regression to associate the sample bicluster with the outcome and obtain the MSE or error rates from the regression model.

\subsection{Simulation Results}

\par
\begin{figure}
  \centering
  \subfloat[][]{\includegraphics[width=\textwidth]{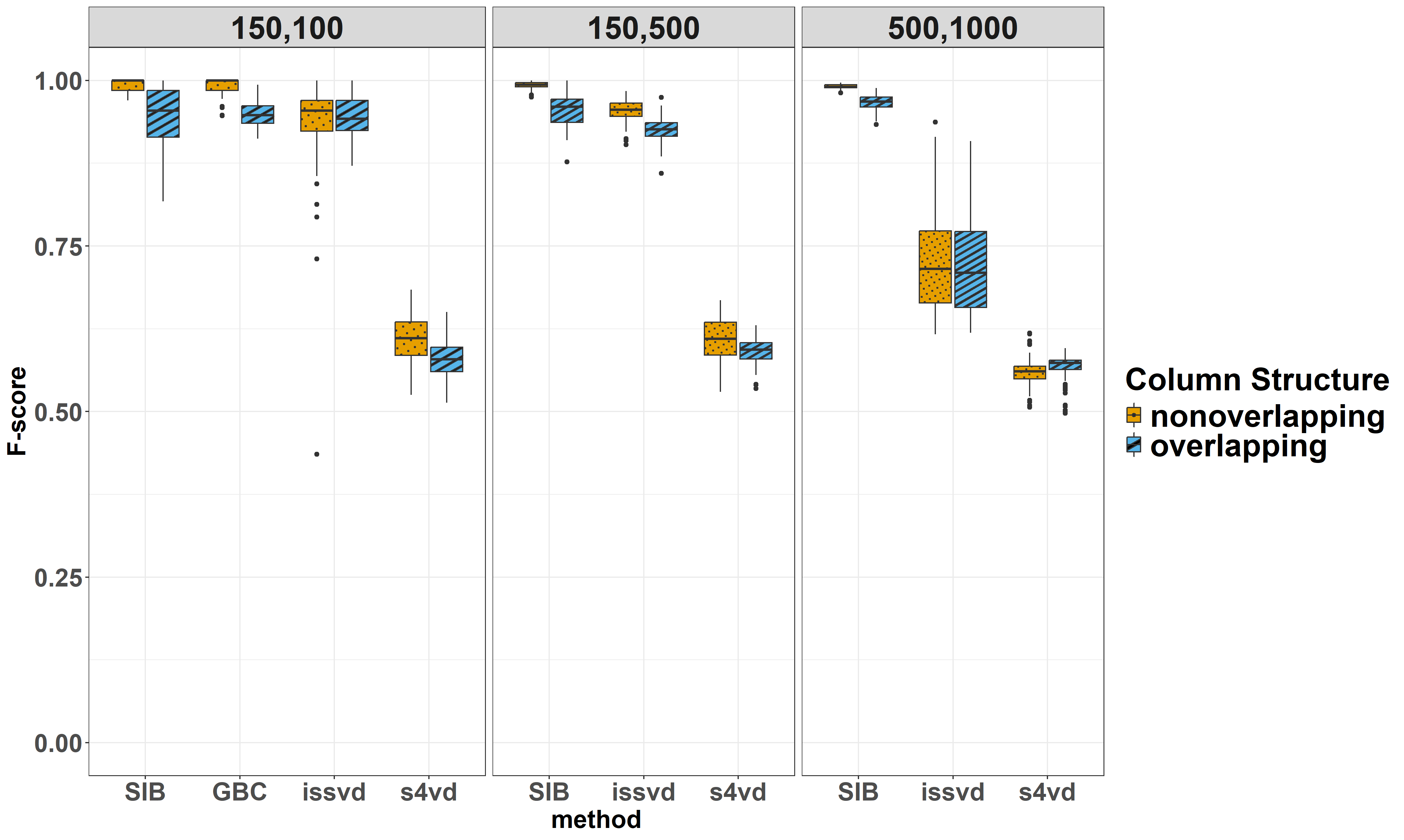}}\\
  \subfloat[][]{\includegraphics[width=\textwidth]{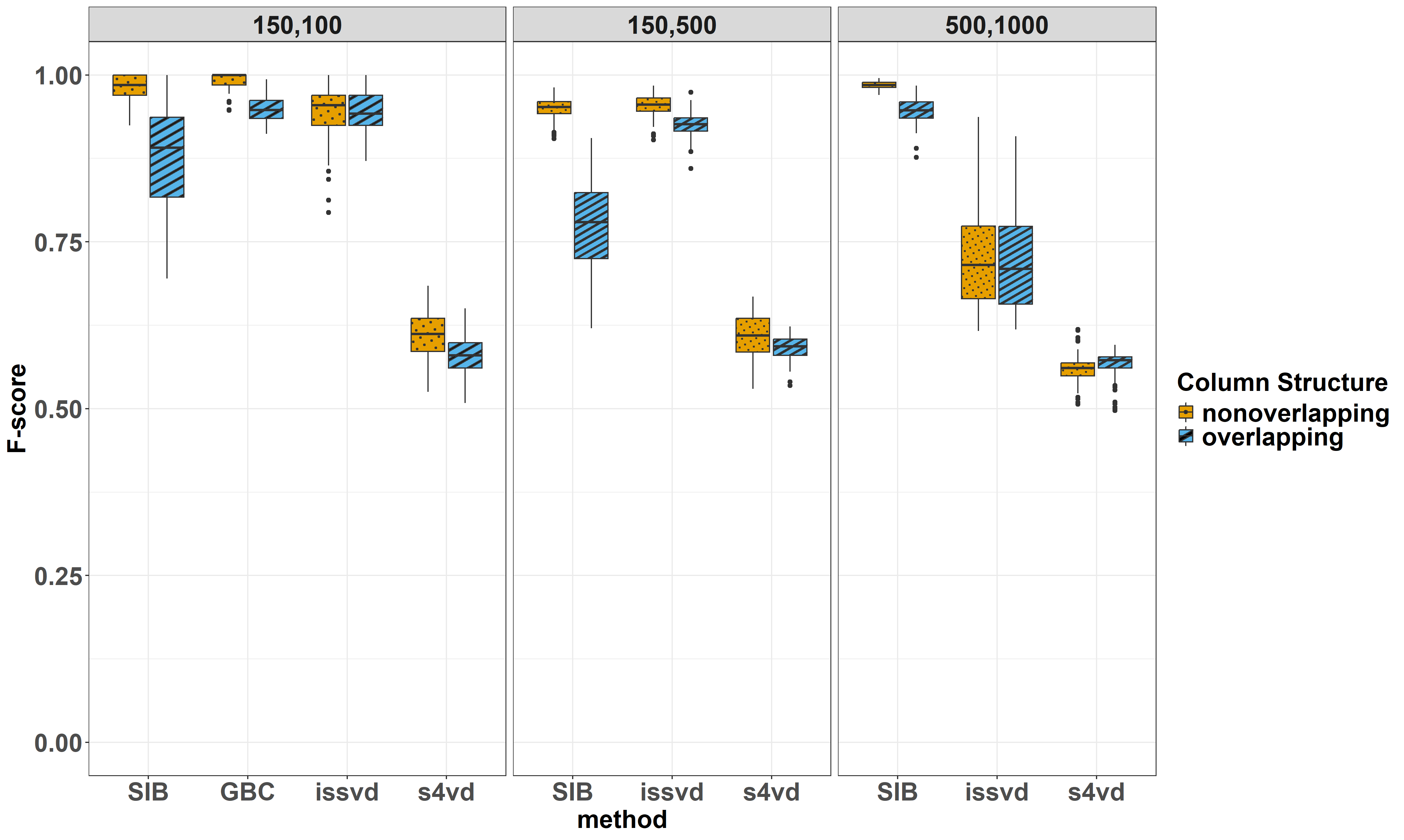}
  }
  \caption{Boxplots of the F-score from 100 reptitions for all methods when the outcome is (a) continuous and (b) binary. The title indicates the number of samples and variables per view (e.g. 150,100 means that the plot for the simulation setting with $n=150$ samples and $p^{(d)}=100$ variables per view $d$). The color of the boxes indicates whether or not we assume the columns to be nonoverlapping when estimating the biclusters.}
  \label{fig:sim_res_F}
\end{figure}

 The simulation results for all settings are summarized in the following tables and figures. In Tables \ref{tab:sim_gg} and \ref{tab:sim_gb}, we show all the evaluation metrics defined in Section \ref{ss:evaluation} for continuous outcome and binary outcome, respectively. Figure \ref{fig:sim_res_F} shows the F-score of these methods in different sets of sample sizes, different numbers of variables, and different types of outcomes. We choose to only show a boxplot of the F-score and do not include the boxplots of the relevance and recovery because in our simulation studies, most pairs of relevance and recovery are close to each other, and the F-score is a good summary for them. Additional figures such as the boxplot of the training and testing MSE are presented in the supporting information Section B.4. In our simulation, we emphasize the performance when we make the correct assumption that the columns are not overlapping, and we use the results under the overlapping assumption as a sensitivity analysis to evaluate the robustness of our method under a weaker assumption.

\begin{table}
\caption{Simulation results when the outcome is a continuous variable that follows a Gaussian distribution. "NOver" denotes whether or not we assume the underlying structure is non-overlapping. "Rec", "Rel", "F-score", "FP", and "FN" are the average recovery, relevance, F-score, false positive, and false negative, respectively, from the training set across all 100 repetitions and are defined in Section \ref{ss:evaluation}. "MSE" and "MSE test" are the average mean squared errors of the outcome in the training and testing sets, respectively.}
\label{tab:sim_gg}
\begin{center}
\begin{tabular}{|c c c c c c c c c c c|} 
 \hline
 $n$&$p$ & Methods &NOver & Rec & Rel & F-score & FP & FN & MSE & MSE test\\
 \hline
 150&100 & SIB & True& \textbf{0.995} & \textbf{0.995} & \textbf{0.995} & \textbf{0.0001} & \textbf{0} & 0.993 & 1.111\\
 150&100 & SIB & False& 0.945 & 0.945 & 0.945 & 0.002 & \textbf{0} & 0.993 & 1.111\\
 150&100 & GBC &True&  0.990 & 0.990 & 0.990 & 0.0002 & \textbf{0} & \textbf{0.982} & 1.870\\
 150&100 & GBC &False&  0.947 & 0.947 & 0.947 & 0.002 & \textbf{0} & 1.152 & 1.878\\
 150&100 & iSSVD &True&  0.942 & 0.934 & 0.936 & 0.018 & 0.039 & 1.039 & 1.027\\
 150&100 & iSSVD &False&  0.942 & 0.942 & 0.942 & 0.022 & \textbf{0} & 0.997 & 1.000\\
 150&100 & S4VD &True&  0.764 & 0.509 & 0.611 & 0.010 & 0.001 & 1.085 & 1.214\\
 150&100 & S4VD &False&  0.723 & 0.482 & 0.579 & 0.013 & 0.0003 & 1.115 & \textbf{0.999}\\
 \hline
 150&500 & SIB & True& \textbf{0.973} & \textbf{0.973} & \textbf{0.973} & \textbf{0.0002} & 0.0007 & \textbf{0.996} & 1.052\\
 150&500 & SIB & False& 0.936 & 0.936 & 0.936 & 0.002& 0.0007 & 0.996 & 1.052 \\
 150&500 & iSSVD &True& 0.954  & 0.954 & 0.954 & 0.014 & 0.003 & 0.996 & 1.016\\
 150&500 & iSSVD &False&  0.925 & 0.925  & 0.925 & 0.027 & \textbf{0} & 0.998 &\textbf{1.014}\\
 150&500 & S4VD &True&  0.762 & 0.508 & 0.609 & 0.009 & 0.001 & 1.104 & 1.037\\
 150&500 & S4VD &False&  0.738 & 0.492 & 0.591 & 0.012 & 0.0003 & 1.101 & 1.037\\
 \hline 
 500&1000 & SIB & True& \textbf{0.991} & \textbf{0.991} & \textbf{0.991} & 0.0003 & 0 & \textbf{1.006} & 1.104\\
 500&1000 & SIB & False& 0.966 & 0.966 & 0.966 & 0.001 & \textbf{0} & \textbf{1.006} & 1.104\\
 500&1000 & iSSVD & True& 0.724 & 0.724 & 0.724 & 0.066 & 0.046 & 1.494 & 1.173\\
 500&1000 & iSSVD & False& 0.725 & 0.725 & 0.725 & 0.078 & 0.042 & 1.502 & 1.176\\
 500&1000 & S4VD & True& 0.699 & 0.466 & 0.559 & 0.014 & 0.0008 & 1.037 & \textbf{1.014}\\
 500&1000 & S4VD & False& 0.705 & 0.470 & 0.564 & 0.015 & 0.0001 & 1.034 & 1.015\\
 \hline
\end{tabular}
\end{center}
\end{table}

\begin{table}
\caption{Simulation results when the outcome is a binary variable that follows a Bernoulli distribution. "NOver" denotes whether or not we assume the underlying structure is non-overlapping. "Rec", "Rel", "F-score", "FP", and "FN" are the average recovery, relevance, F-score, false positive, and false negative, respectively, from the training set across all 100 repetitions and are defined in Section \ref{ss:evaluation}. "ER" and "ER test" are the average classification error rates in the training and testing sets, respectively.}
\label{tab:sim_gb}
\begin{center}
\begin{tabular}{|c c c c c c c c c c c|} 
 \hline
 $n$&$p$ & Methods &NOver & Rec & Rel & F-score & FP & FN & ER & ER test\\
 \hline
 150&100 & SIB & True& 0.967 & 0.967 & 0.967 & 0.0006 & 0.0006 & 0.269 & 0.258\\
 150&100 & SIB & False& 0.861 & 0.861 & 0.861 & 0.006 & 0.0006 & 0.269 & 0.258\\
 150&100 & GBC &True& \textbf{0.990} & \textbf{0.990} & \textbf{0.990} & \textbf{0.0002} & \textbf{0} & 0.271 & 0.249\\
 150&100 & GBC &False&  0.947 & 0.947 & 0.947 & 0.002 & \textbf{0} & 0.270 & 0.254\\
 150&100 & iSSVD &True&  0.942 & 0.934 & 0.936 & 0.018 & 0.039 & 0.269 & 0.251\\
 150&100 & iSSVD &False&  0.942 & 0.942 & 0.942 & 0.022 & \textbf{0} & 0.269 & \textbf{0.249}\\
 150&100 & S4VD &True&  0.764 & 0.509 & 0.611 & 0.010 & 0.001 & \textbf{0.267} & 0.250\\
 150&100 & S4VD &False&  0.723 & 0.482 & 0.579 & 0.013 & 0.0003& 0.269 & 0.254\\
 \hline
 150&500 & SIB & True& 0.949 & 0.949 & 0.949 & \textbf{0.002} & \textbf{0} & 0.273 & 0.294\\
 150&500 & SIB & False& 0.773 & 0.773 & 0.773 & 0.010& \textbf{0} & 0.273 & 0.294 \\
 150&500 & iSSVD &True& \textbf{0.954}  & \textbf{0.954} & \textbf{0.954} & 0.014 & 0.003 & 0.273 & \textbf{0.239}\\
 150&500 & iSSVD &False&  0.925 & 0.925  & 0.925 & 0.027 & \textbf{0} & 0.273 &\textbf{0.239}\\
 150&500 & S4VD &True&  0.762 & 0.508 & 0.609 & 0.009 & 0.001 & 0.273 & 0.291\\
 150&500 & S4VD &False&  0.738 & 0.492 & 0.591 & 0.012 & 0.0003 & \textbf{0.270} & 0.290\\
 \hline 
 500&1000 & SIB & True& \textbf{0.985} & \textbf{0.985} & \textbf{0.985} & 0.002 & \textbf{0} & \textbf{0.268} & 0.280\\
 500&1000 & SIB & False& 0.946 & 0.946 & 0.946 & \textbf{0.0006} & \textbf{0} & \textbf{0.268} & 0.280\\
 500&1000 & iSSVD & True& 0.724 & 0.724 & 0.724 & 0.066 & 0.046 & 0.278 & 0.280\\
 500&1000 & iSSVD & False& 0.725 & 0.725 & 0.725 & 0.078 & 0.042 & 0.279 & 0.280\\
 500&1000 & S4VD & True& 0.699 & 0.466 & 0.559 & 0.014 & 0.0008 & \textbf{0.268} & 0.280\\
 500&1000 & S4VD & False& 0.705 & 0.470 & 0.564 & 0.015 & 0.0001 & 0.271 & \textbf{0.279}\\
 \hline
\end{tabular}
\end{center}
\end{table}

\subsection{Discussion of Simulation Results}
\par  We focus on the accuracy of the biclusters in the main text and leave the outcome prediction results in the supporting information Section B.4 since the performance of the outcome prediction is similar across all methods. Table~\ref{tab:sim_gg} shows that our proposed method has the best average recovery, relevance, and F-1 score for all the sample sizes and numbers of variables when the outcomes are continuous. Although GBC shows a comparable result compared to ours when the sample size is 150 and the number of variables is 100, it requires intensive computational time, making it impractical for larger settings. The computational time of all the methods is summarized in Web Table B.1 in the supporting information. Both iSSVD and S4VD have worse performance than ours. Specifically, S4VD struggles to identify the third bicluster, which is treated as a background bicluster by S4VD, leading to a worse performance compared to all other methods. Additionally, Figure \ref{fig:sim_res_F} (a) highlights that our proposed method has the smallest variance under all scenarios and assumptions.

\par Our proposed method, GBC, and S4VD are all sensitive to the nonoverlapping assumption. Both Table \ref{tab:sim_gg} and Figure \ref{fig:sim_res_F} (a) show that the performance of these methods declines if we allow the columns to overlap, which matches our expectation given that the underlying assumption of the bicluster structure is nonoverlapping. The false positive rates increase, and the false negative rates decrease when we remove the nonoverlapping assumption.  

\par The number of variables and the sample size also affect the accuracy of the biclustering results. We notice a slight decrease in recovery, relevance, and F-score for our proposed method when the number of variables is increased to 500, while the performance of iSSVD and S4VD is not significantly impacted. These metrics even increase a little for iSSVD when we assumed a nonoverlapping structure for columns and for S4VD when we assumed an overlapping structure for columns. Despite a slight decline in performance, our method still outperforms the comparison methods in all metrics for biclustering accuracy. When both the sample size and the number of variables increase, our proposed method achieves a recovery, relevance, and F-score close to 1, and a false positive and false negative close to 0. In contrast, all the compparison methods exhibit a decline in biclustering accuracy.

\par The performance of the comparison methods does not change when the outcome is binary, since these methods do not account for the outcome in their estimation. In Table~\ref{tab:sim_gb} and Figure~\ref{fig:sim_res_F} (b), we observe a slight decrease in the performance of our proposed method when the outcome is binary, but our method still performs well when the nonoverlapping structure is correctly specified. The decrease in performance is expected, as the number of biclusters exceeds the number of outcome classes \citep{wangSupervisedConvexClustering2023}. Under this generation scheme, correctly assigning samples to the bicluster with a balanced mixture of positive and negative cases becomes more challenging. The impact of more variables and larger sample sizes is similar to that when the outcome is continuous.

\par Table~\ref{tab:sim_gg} and Table~\ref{tab:sim_gb} summarize the performance of fitting and predicting outcomes with the proposed and comparison methods. The results show that the performance is similar across all the methods in most cases, suggesting that our method performs well in predicting the biclusters and the outcome for new samples. Boxplots of training and testing performances for the outcomes are included in the supporting information Section B.4.

\par The simulation results highlight the superiority of our proposed method compared to the other methods. Our method outperforms the comparison methods in almost all cases when correctly assuming a nonoverlapping column structure. The performance of the proposed method decreases slightly as the number of variables increases, while it improves when the sample size increases, contrary to the comparison methods. This suggests that our method is suitable for finding biclusters from datasets with various sizes, especially larger datasets.

\section{Real Data Analysis}
\label{s:rda}

\subsection{Background}

\par In this section, our objective is to integrate lipidomics and imaging data from the ADNI study to identify subgroups of AD with varying progression to better understand the heterogeneity in AD. Because we use the outcome, MMSE, in the detection of our biclusters, we expect to detect biclusters that are significantly different in their AD risk and progression compared to existing methods that do not use any outcome. The imaging data are summary measures (e.g., volume, surface areas, thickness, etc.) from various brain regions obtained from magnetic resonance imaging (MRI). Imaging and lipidomics data are obtained from $n=480$ individuals, of which 177 are cognitively normal (CN), 235 have mild cognitive impairment (MCI), and 68 have AD. There are 337 imaging variables and 768 lipids, respectively. We use MMSE, a widely used cognitive score, as the supervising variable for our proposed method. The MMSE consists of 11 questions, with a total possible score of 30. A higher score on the MMSE indicates better cognitive function\citep{kurlowiczMiniMentalStateExamination1999}. Our analytical data are $\bm{X}^{(1)}$, $\bm{X}^{(2)}$, and $\bm{y}$, for lipidomics, imaging, and MMSE, respectively.

\par We preprocessed the data before applying the biclustering methods. We applied unsupervised filtering to both the lipidomics and imaging data, removing lipids and imaging variables with variances below the 5th percentile. We log-transformed the lipidomics data to reduce skewness, but we chose not to log-transform the imaging variables to preserve their interpretability. We standardized each lipid and imaging variable to have a mean of zero and a variance of one, ensuring that all variables were on the same scale. In implementing our proposed method, we fixed the number of biclusters, $K=3$, to represent the three AD groups in our data (i.e., CN, MCI, and AD). We assumed a nonoverlapping column structure for all methods to enhance interpretability. The tuning parameters for iSSVD and S4VD were selected based on the authors' recommendations. When applying S4VD, we concatenated the lipidomics and imaging data, as S4VD is only applicable to a single view. We used baseline variables (e.g., demographics), to characterize the identified biclusters.

\subsection{Results}

\par
\begin{figure}
    
    \centering
    \begin{minipage}{\textwidth}
        \centering
        \includegraphics[width=\textwidth]{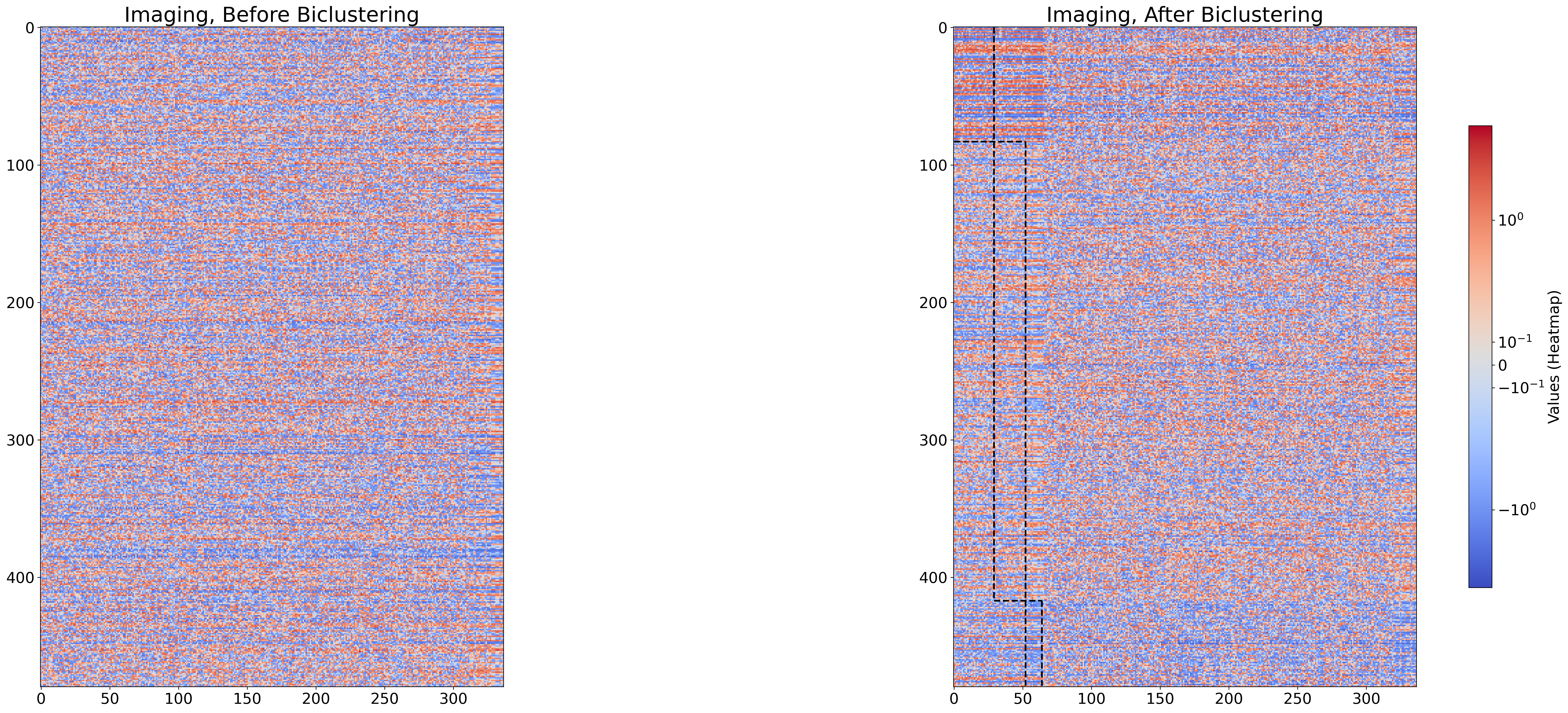}
        \subcaption{}
    \end{minipage}

    \vspace{0.5cm} 

    \begin{minipage}{\textwidth}
        \centering
        \includegraphics[width=\textwidth]{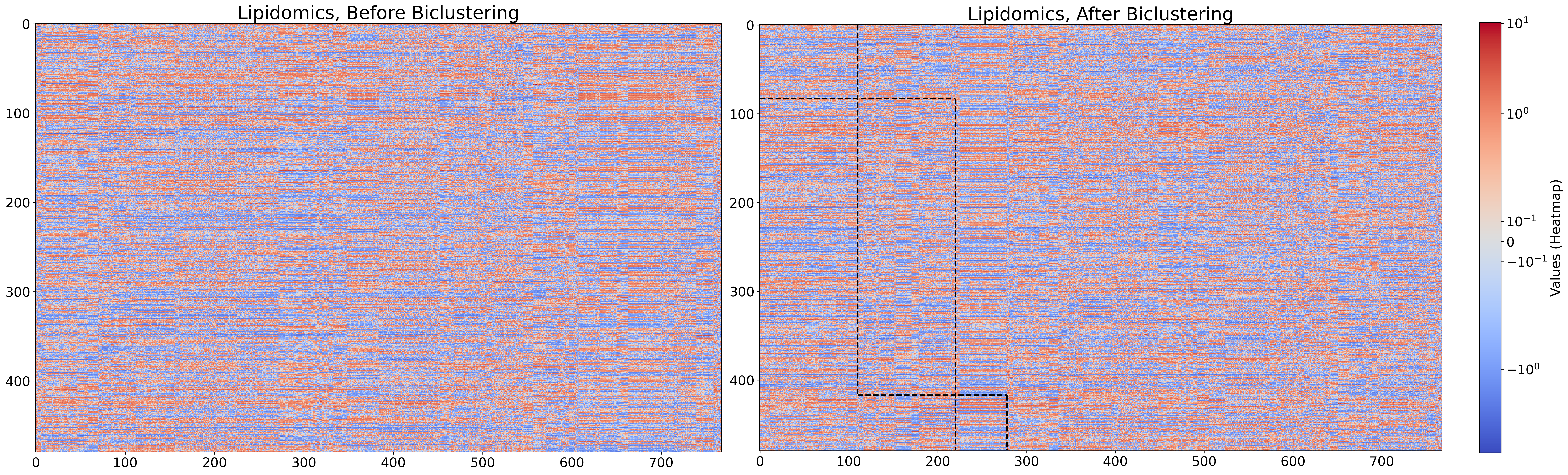}
        \subcaption{}
    \end{minipage}
    
    \vspace{0.5cm} 
    
    \begin{minipage}{\textwidth}
        \centering
        \includegraphics[width=.8\textwidth]{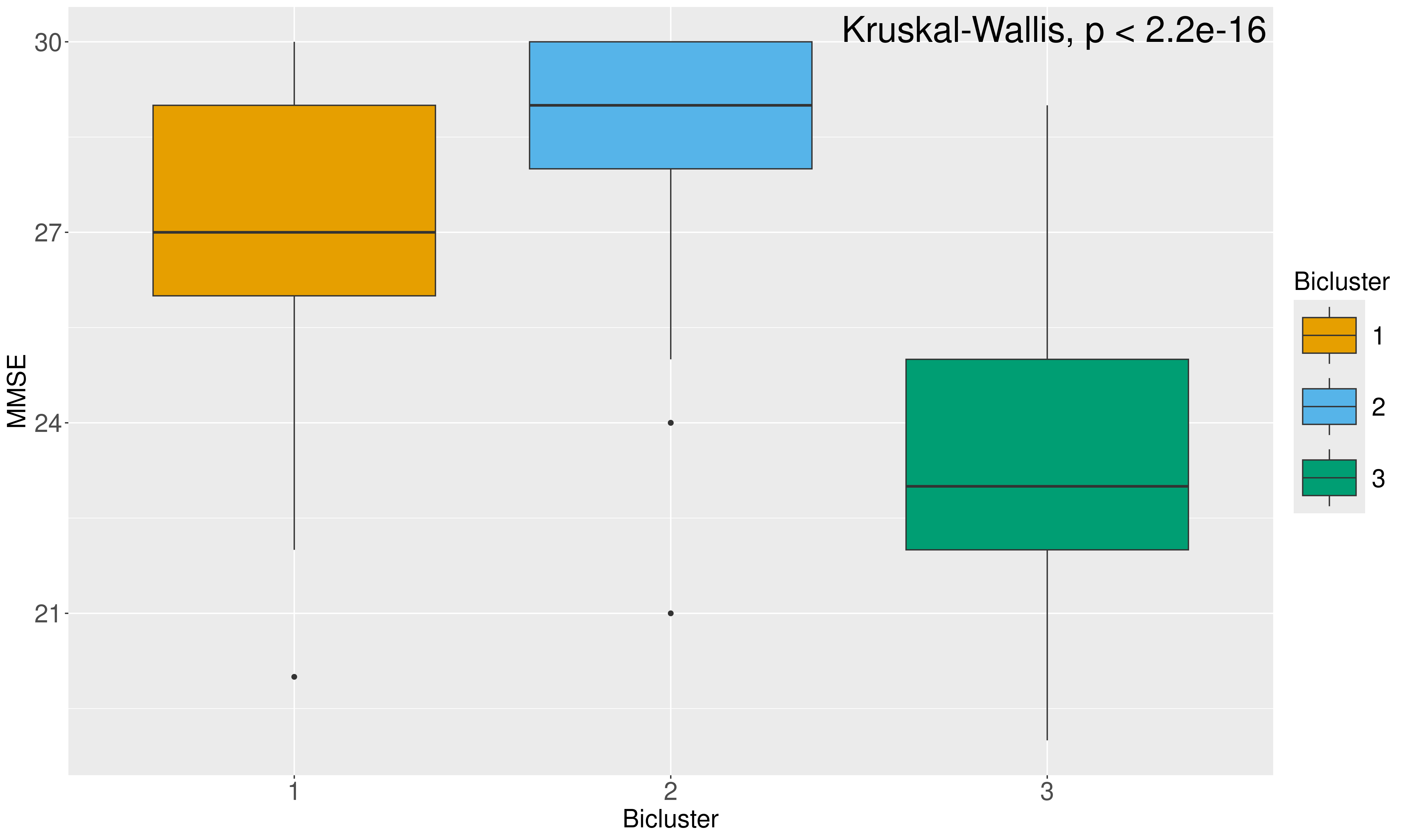}
        \subcaption{}
    \end{minipage}
\caption{Visualization of the results: heatmaps of (a) imaging variables and (b) lipidomics, and (c) boxplot of the Mini Mental State Examination (MMSE). The heatmaps on the left are plotted using the original dataset, and the ones on the right are reordered using the biclustering results. The colors on the heatmaps are on a log scale for better readability. Biclusters are labeled using dashed lines.}
\label{fig:rda_res}
\end{figure}

 Our method identifies three biclusters, consisting of 84, 334, and 62 individuals (samples), respectively. In the first bicluster, we identify 111 lipids and 30 imaging variables; in the second, 99 lipids and 23 imaging variables; and in the third, 69 lipids and 12 imaging variables. A list of the variable names for these identified lipids and imaging variables can be found in Web Tables C.2 and C.3. Table \ref{tab:rda} characterizes the biclusters identified by our method based on some key phenotypic variables.  A full table with all variables considered is included in Web Table C.1 in the supporting information. The estimated mean MMSE $\bm{\widehat{\beta}}$ from our method is $(26.95, 28.64, 23.51)$ for the three biclusters, respectively. This suggests that cognitive function is best in the second bicluster, followed by the first bicluster, with the third bicluster showing the lowest cognitive function. Additionally, we can infer that the MMSE score for an individual in the second bicluster is expected to be approximately 1.5 points higher than that of an individual in the first bicluster, and about 5 points higher than that of an individual in the third bicluster. Figures \ref{fig:rda_res}(a) and (b) display heatmaps of the lipidomics and imaging variables. On the left side of each figure, we present the heatmaps before biclustering, and on the right side, we show the heatmaps after reordering the data based on the biclustering results, allowing for easier visualization. For improved readability, we display the colors on a log-scale and highlight the biclusters using dashed lines. Figure \ref{fig:rda_res} (c) presents a boxplot of the outcome variable, MMSE. iSSVD also discovered three biclusters, but they are not associated with any clinical variable of interest.  S4VD did not discover any stable bicluster. Please refer to supporting information Section C for more details.


\begin{table}[!t]


\begin{center}
\small
    \begin{tabular}{lcccc}
    \toprule
    \textbf{Bicluster} & \textbf{1}  & \textbf{2} & \textbf{3}& \\
    & N = 84 &N = 334 & N = 62 & \textbf{p-value}\\ 
    \midrule
    \textbf{Demographical Information} &  &  &  & \\
    Age & 71 (67, 77) & 72 (68, 77) & 75 (68, 79) & 0.15 \\ 
    Gender &  &  &  & 0.11 \\ 
        Female & 33 (39\%) & 174 (52\%) & 30 (48\%) &  \\ 
        Male & 51 (61\%) & 160 (48\%) & 32 (52\%) &  \\ 
    Ethnicity &  &  &  & 0.3 \\ 
        Hisp/Latino & 0 (0\%) & 11 (3.3\%) & 3 (4.8\%) &  \\ 
        Not Hisp/Latino & 84 (100\%) & 322 (96\%) & 59 (95\%) &  \\ 
        Unknown & 0 (0\%) & 1 (0.3\%) & 0 (0\%) &  \\ 
    Race &  &  &  & 0.4 \\ 
        Am Indian/Alaskan & 0 (0\%) & 1 (0.3\%) & 0 (0\%) &  \\ 
        Asian & 1 (1.2\%) & 6 (1.8\%) & 0 (0\%) &  \\ 
        Black & 4 (4.8\%) & 15 (4.5\%) & 4 (6.5\%) &  \\ 
        Hawaiian/Other PI & 1 (1.2\%) & 0 (0\%) & 0 (0\%) &  \\ 
        More than one & 1 (1.2\%) & 2 (0.6\%) & 0 (0\%) &  \\ 
        Unknown & 0 (0\%) & 0 (0\%) & 1 (1.6\%) &  \\ 
        White & 77 (92\%) & 310 (93\%) & 57 (92\%) &  \\ 
    Education & 16.00 (14.00, 18.00) & 17.00 (14.00, 18.00) & 16.00 (14.00, 18.00) & 0.044 \\ 
    
    Dropped activities and interests &  &  &  & 0.044 \\ 
        No(0) & 70 (83\%) & 305 (91\%) & 52 (84\%) &  \\ 
        Yes(1) & 14 (17\%) & 29 (8.7\%) & 10 (16\%) &  \\ 
    Alcohol abuse & 3 (3.6\%) & 10 (3.0\%) & 7 (11\%) & 0.016 \\ 
    \midrule
    \textbf{AD Related Conditions} &&&&\\
    Family History of AD & 26 (31\%) & 126 (38\%) & 19 (31\%) & 0.3 \\ 
    Diagnosis (3-level) &  &  &  & $<$0.001 \\ 
        CN & 22 (26\%) & 150 (45\%) & 5 (8.1\%) &  \\ 
        MCI & 47 (56\%) & 170 (51\%) & 18 (29\%) &  \\
        AD & 15 (18\%) & 14 (4.2\%) & 39 (63\%) &  \\ 
    
    MMSE & 27 (26, 29) & 29 (28, 30) & 23 (22, 25) & $<$0.001 \\ 
    CDRSB & 1.50 (0.00, 2.50) & 0.50 (0.00, 1.50) & 3.50 (2.00, 5.00) & $<$ 0.001 \\ 
    ADAS13 & 17 (11, 22) & 11 (7, 16) & 27 (20, 33) & $<$0.001 \\ 
        Unknown & 0 & 1 & 1 &  \\ 
    mPACCdigit & -7 (-11, -3) & -2 (-5, 0) & -16 (-19, -12) & $<$0.001 \\ 
    mPACCtrailsB & -6.2 (-9.1, -1.6) & -1.7 (-4.6, 0.6) & -14.3 (-16.1, -11.1) & $<$0.001 \\ 
    \midrule
    \textbf{AD Biomarkers} &&&&\\
    Hippocampus & 6,970 (5,903, 8,054) & 7,291 (6,679, 7,864) & 6,030 (5,111, 6,702) & $<$0.001 \\ 
        Unknown & 8 & 31 & 5 &  \\ 
    APOE4 &  &  &  & $<$0.001 \\ 
        0 & 35 (42\%) & 201 (60\%) & 22 (35\%) &  \\ 
        1 & 36 (43\%) & 108 (32\%) & 28 (45\%) &  \\ 
        2 & 13 (15\%) & 25 (7.5\%) & 12 (19\%) &  \\ 
    \bottomrule
\end{tabular}
\end{center}
\caption{Key baseline characteristics of ADNI participants by the biclusters identified with the proposed method. Values are median (first and third quartiles) for continuous variables, and N (percentages) for binary/categorical variables. P-values are calculated by comparing all three obtained biclusters using the Kruskal-Wallis test for continuous variables, and the Chi-square test or Fisher's exact test for binary/categorical variables depending on the sizes of the cells.}
\label{tab:rda}
\end{table}

\subsection{Discussion of Real Data Results}
\subsubsection{Cognitive Functions are Significantly Different among Identified Biclusters}
\par In this subsection, we focus on our proposed method, as the other two methods did not identify any clinically meaningful biclusters. We begin by examining the outcome variable, the MMSE scores, across the biclusters identified by our proposed method.
Table \ref{tab:rda} shows that the MMSE score is highest in bicluster 2, followed by bicluster 1. The median MMSE score is lowest in bicluster 3. The boxplot of MMSE scores in Figure \ref{fig:rda_res} (c) further confirms this observation. The second bicluster consists mainly of individuals with high MMSE scores, while the third bicluster is composed mainly of individuals with low MMSE scores. This suggests that bicluster 2 is characterized by good cognitive function with no signs of impairment, bicluster 1 is associated with mild cognitive decline, and bicluster 3 reflects a more significant loss of cognitive function. The diagnosis variable (e.g., AD status) supports this interpretation: the second bicluster has the highest proportion of CN individuals ($45\%$), with almost no participants diagnosed with AD. The third bicluster, on the other hand, includes the highest proportion of individuals with AD ($63\%$) and nearly no CN individuals. The first bicluster has the highest proportion of individuals with MCI ($56\%$), along with a small proportion of individuals with CN or AD. Furthermore, the identified biclusters are significantly associated with other cognitive scores, including the Clinical Dementia Rating - Sum of Boxes (CDR-SB), the Alzheimer's Disease Assessment Scale - Cognitive (ADAS13), and the Preclinical Alzheimer Cognitive Composite (mPACC digit and mPACC trailsB).

\par In addition to covariates directly related to AD, we examined associations between the identified biclusters and other baseline variables to further characterize the biclusters. The biclusters are not significantly associated with most demographic factors, such as age, gender, race, or ethnicity. However, some social determinants of health are found to differ between the three biclusters, including education level, dropped activities and interests, and alcohol abuse. These findings align with the existing literature that lower education levels and fewer leisure activities are associated with a higher risk of developing AD \citep{zhangEpidemiologyAlzheimersDisease2021}. No significant associations are found between clinical comorbidities and the biclusters. A summary of these associations can be found in Web Table C.4 in the supporting information.

\subsubsection{Brain Regions and Lipidomics Categories Show Potential in Characterizing Cognitive Functions}

\par Figure \ref{fig:rda_res} (a) and (b) present heatmaps for visualizing our biclustering results. Figure \ref{fig:rda_res} (a) shows the heatmaps before and after reordering the rows and columns of the imaging data using the biclustering results. There is no noticeable pattern before reordering. After reordering, we notice a clear distinct difference between the variables selected by these three biclusters and the other variables, indicating that the proposed method effectively identifies the truly important variables. Additionally, the samples in each bicluster display a notable difference in these biclusters. The samples in the first bicluster exhibit relatively higher values compared to the other two biclusters, while the samples in the third bicluster have much lower values in all variables compared to the other two. Similarly, in Figure \ref{fig:rda_res}(b), the heatmap of the original lipidomics data on the left shows no clear patterns, and the underlying structure becomes more apparent after reordering. The top left corner of the reordered heatmap represents the first bicluster, and we observe that the individuals in the first bicluster exhibit lower expressions of these selected lipids than the other two biclusters, particularly compared to those in the third bicluster. The block at the bottom right represents the third bicluster, where a distinct pattern is observed among the selected lipids, differentiating them from the other lipids across all samples.

\par
\begin{figure}
        \centering
        \begin{subfigure}[b]{0.475\textwidth}
            \centering
            \includegraphics[width=\textwidth]{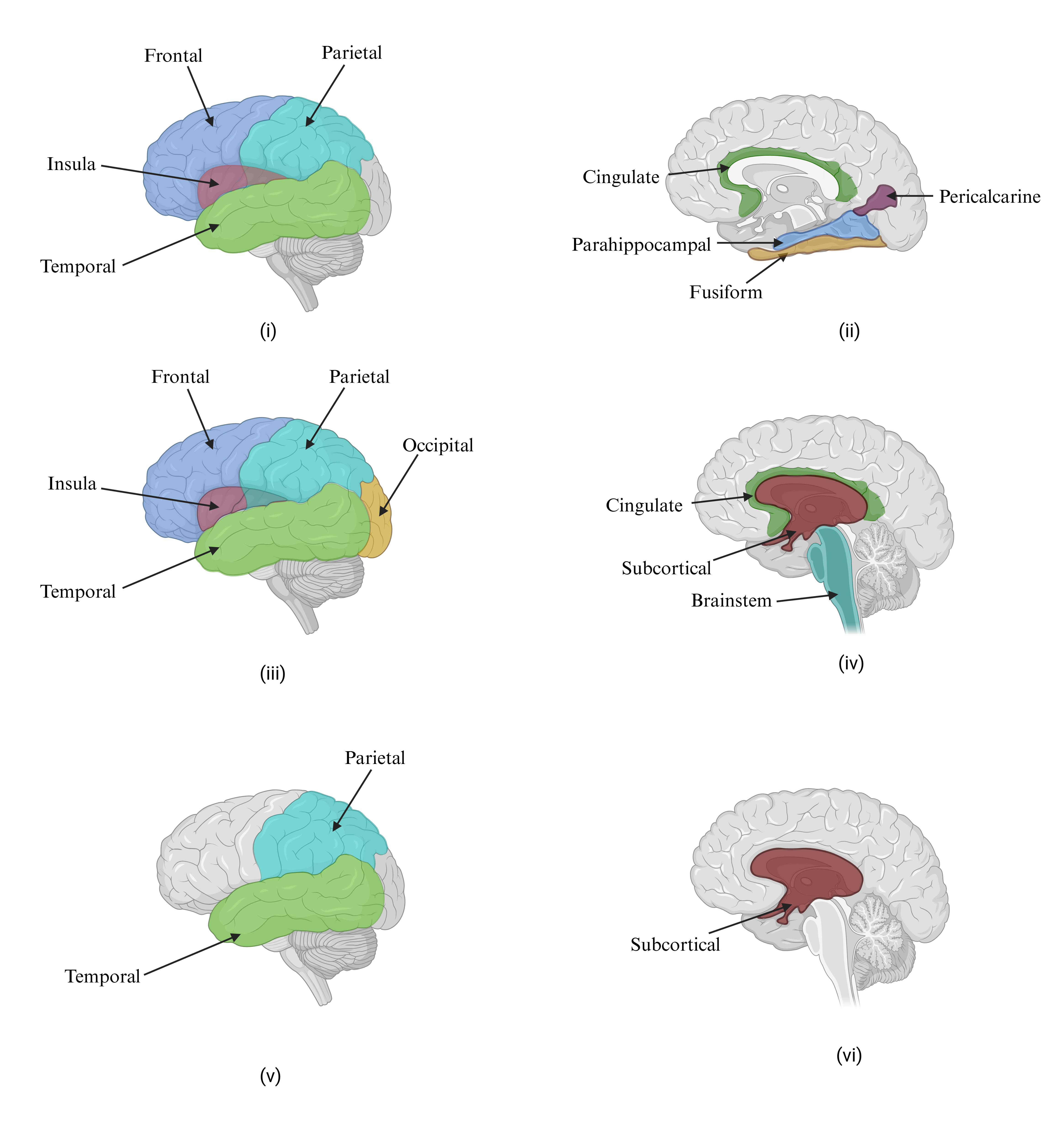}
            \caption{}
        \end{subfigure}
        \hfill
        \begin{subfigure}[b]{0.475\textwidth}  
            \centering 
            \includegraphics[width=\textwidth]{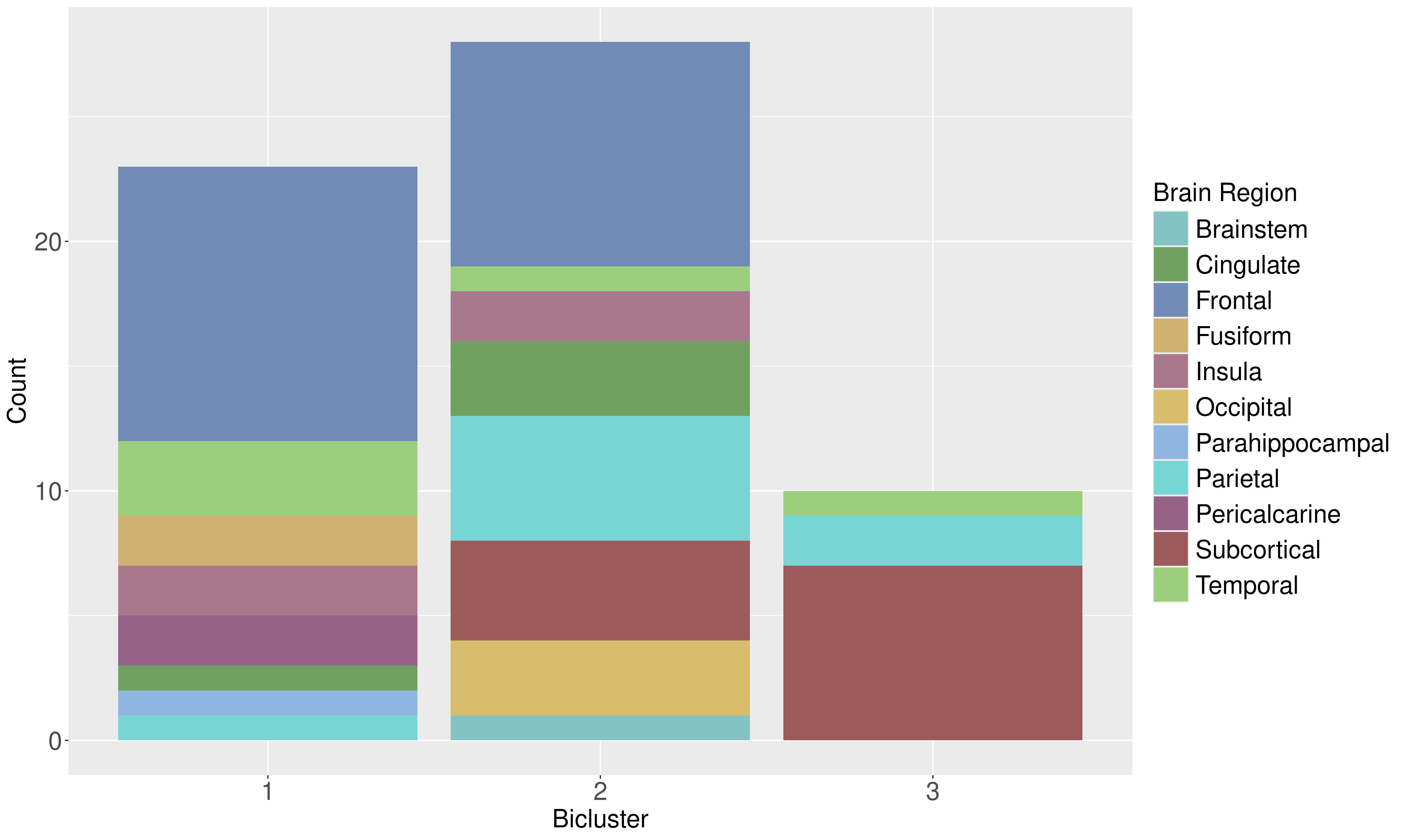}
            \caption{}
        \end{subfigure}
        \vskip\baselineskip
        \begin{subfigure}[b]{0.475\textwidth}   
            \centering 
            \includegraphics[width=\textwidth]{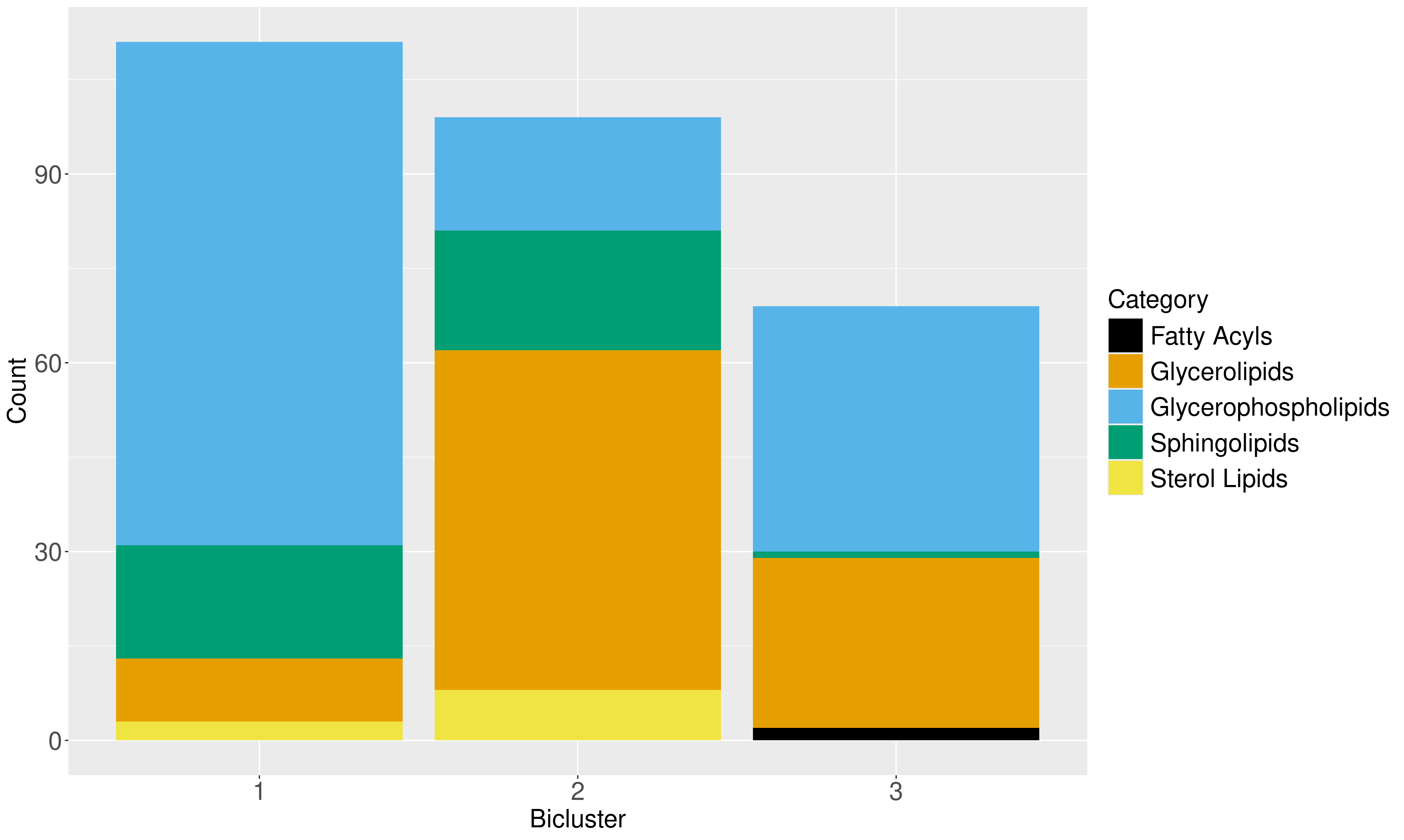}
            \caption{}
        \end{subfigure}
        \hfill
        \caption{Visualizations of region and category analyses. (a) shows the regions of brains that are selected in each bicluster. (i), (iii), and (v) shows the regions selected by the first, second, and third bicluster in lateral view, and (ii), (iv), and (vi) shows these regions in sagittal view. (b) and (c) are barplots of the number of variables from each class by each bicluster in each dataset. (b) shows the number of lipids in each LIPID MAPS category by biclusters; (c) shows the number of imaging varaibles in each brain reigion by biclusters.} 
        \label{fig:barplots}
\end{figure}

 To obtain further understanding of the image and lipid characteristics for the biclusters, we categorized the imaging variables for each bicluster by brain regions and the lipid variables by lipid categories using LIPID MAPS \citep{liebischUpdateLIPIDMAPS2020}. Figure \ref{fig:barplots} (a) and (b) show the brain regions selected for each bicluster. Figure \ref{fig:barplots} (a) presents both lateral and sagittal views for each bicluster to allow regions inside brain to be visible. We notice that parietal and temporal regions are involved in all three biclusters, and frontal, insula, and cingulate regions are selected in both the first and the second biclusters. The first bicluster, which corresponds to mildly cognitively impaired participants, is characterized by parahippocampal, fusiform, and pericalcarine regions. These regions have been linked to MCI in current literatures. For example, parahippocampal can be used to predict conversion to AD among MCI patients \citep{devanandMRIHippocampalEntorhinal2012}, fusiform is shown to be altered among patients with MCI \citep{caiAlteredFunctionalConnectivity2015}, and pericaclarine is important in classifying subtypes of MCI \citep{guanClassifyingMCISubtypes2017}. The second bicluster, corresponding to the group with best cognitive functions, is characterized by occipital and brainstem. The third bicluster, which contains the majority cognitively impaired participants, is characterized by subcortical region. Furthermore, most of these imaging variables in subcortical areas are related to gray matter and white matter, which are also strongly linked to AD \citep{vandemortelGreyMatterLoss2021,sachdevAlzheimersDiseaseWhite2013}

\par Our analysis of the lipidomics categories also identified glycerophospholipids and glycerophospholipids as promising categories for further studies. Figure \ref{fig:barplots} (c) shows the number of lipids characterizing each bicluster. We notice that the first bicluster is characterized by glycerophospholipids, the second bicluster is characterized by glycerolipids, and the third bicluster consists of similar numbers of lipids from these two categories. Both glycerophospholipids and glycerolipids are shown to be pertubed among AD patients \citep{akyolLipidProfilingAlzheimers2021}, and alteration in glycerophospholipids may be a major determinant of cognitive loss, supporting its selection in the first and the third bicluster \citep{woodLipidomicsAlzheimersDisease2012}.



\par The real data analysis highlights the good performance of our proposed method compared to the other methods.  The biclusters detected by our method are significantly associated with clinical variables related to AD and also differed in terms of social determinants of health. In contrast, none of the comparison methods finds interpretable biclusters. Furthermore, the selected lipids and imaging variables are validated by existing literature, providing scientifical interpretation of our results. These results suggest that our proposed method can provide insights into the heterogeneity of complex diseases such as AD by identifying meaningful subgroups and their characteristics.

\section{Conclusion}

\par In this paper, we introduced a novel method, Supervised Integrative Biclustering (SIB), which uses clinical outcome data to guide bicluster detection across multiple data sources. This approach improves the interpretability of the detected biclusters, setting it apart from existing unsupervised biclustering methods for multi-view data that may not yield clinically meaningful findings. Additionally, it accommodates various data distributions, allowing it to analyze a broader range of data. Our method can be applied to various complex disease subtyping problems, such as AD, cancer, etc. We demonstrated the overall superiority of our proposed method when compared to other existing methods through extensive simulations.


\par We further demonstrate the effectiveness of our method using lipidomics, imaging, and phenotypic data from the ADNI Study. Our method successfully identified biclusters that are strongly associated with several clinical variables, including AD status and cognitive test scores. In contrast, biclusters detected by existing methods were not associated with any clinical outcomes. Our identified biclusters are also associated with several baseline information, such as education and alcohol abuse. The three biclusters differed in the degree of cognitive impairment: one group showed minimal cognitive decline, another exhibited mild cognitive decline, and the third demonstrated significant cognitive decline. Further, the lipid categories and brian regions characterizing these biclusters appeared to differ, suggesting variations in AD pathology and highlighting the need for further investigation.

\par Our method also has some limitations. First, it depends on a priori-chosen outcome, which requires subject matter expertise. This may hinder the discovery of novel biclusters linked to less common outcomes. Furthermore, restricting the analysis to just one outcome could overlook the potential to uncover subtypes that are defined by multiple outcomes. Another limitation of our method is that we do not use additional data to directly guide the detection of variable clusters. Although the supervising outcome affects the variable clusters indirectly, a knowledge-guided approach may encourage more interpretable variable clusters. Future researchers may consider extending our method to incorporate a direct supervising approach in variable clustering, such as biological graphs, to further enhance the interpretability of our method. Lastly, our method focuses only on the shared component. It may be worth adding an individual component similar to the formulation in JIVE \citep{lockJointIndividualVariation2013}.

\par In conclusion, we have proposed and developed an integrative multi-view biclustering method that allows various data distributions and leverages an outcome to guide the detection of biclusters, enhancing the interpretation of the biclusters. The results from both simulations and real data analyses are promising, suggesting that researches can effectively apply our method for subtyping various complex diseases.  




\section*{Acknowledgments}

\par Dataset in Section~\ref{s:rda} is obtained from the Alzheimer’s Disease Neuroimaging
Initiative (ADNI) database funded by the Alzheimer's Disease Neuroimaging Initiative (ADNI) (National Institutes of Health Grant U01 AG024904) and DOD ADNI (Department of Defense award number W81XWH-12-2-0012). ADNI is funded by the National Institute on Aging, the National Institute of Biomedical Imaging and Bioengineering, and through generous contributions from the following: AbbVie, Alzheimer’s Association; Alzheimer’s Drug Discovery Foundation; Araclon Biotech; BioClinica, Inc.; Biogen; Bristol-Myers Squibb Company; CereSpir, Inc.; Cogstate; Eisai Inc.; Elan Pharmaceuticals, Inc.; Eli Lilly and Company; EuroImmun; F. Hoffmann-La Roche Ltd and its affiliated company Genentech, Inc.; Fujirebio; GE Healthcare; IXICO Ltd.; Janssen Alzheimer Immunotherapy Research \& Development, LLC.; Johnson \& Johnson Pharmaceutical Research \& Development LLC.; Lumosity; Lundbeck; Merck \& Co., Inc.; Meso Scale Diagnostics, LLC.; NeuroRx Research; Neurotrack Technologies; Novartis Pharmaceuticals Corporation; Pfizer Inc.; Piramal Imaging; Servier; Takeda Pharmaceutical Company; and Transition Therapeutics. The Canadian Institutes of Health Research is providing funds to support ADNI clinical sites in Canada. Private sector contributions are facilitated by the Foundation for the National Institutes of Health (www.fnih.org). The grantee organization is the Northern California Institute for Research and Education, and the study is coordinated by the Alzheimer’s Therapeutic Research Institute at the University of Southern California. ADNI data are disseminated by the Laboratory for Neuro Imaging at the University of Southern California. \vspace*{-8pt}

\par \section*{Funding}
Sandra E. Safo and Thierry Chekouo were partially supported by grant \#1R35GM142695 and 1R35GM150537, respectively, of the National Institute of General Medical Sciences of the National Institutes of Health. The content is solely the responsibility of the authors and does not represent the official views of the funding agencies.

\section*{Software}
The SIB package for implementing the method proposed, including simulations, is found at \url{https://github.com/KevenY/SupervisedIntegrativeBiclustering}.


%
\bibliographystyle{unsrtnat} 
\bibliography{ref_final}

\begin{thebibliography}{32}
\providecommand{\natexlab}[1]{#1}
\providecommand{\url}[1]{\texttt{#1}}
\expandafter\ifx\csname urlstyle\endcsname\relax
  \providecommand{\doi}[1]{doi: #1}\else
  \providecommand{\doi}{doi: \begingroup \urlstyle{rm}\Url}\fi

\bibitem[Alzheimer's-Association(2024)]{alzheimersassociation2024AlzheimersDisease2024}
Alzheimer's-Association.
\newblock 2024 alzheimer's disease facts and figures.
\newblock \emph{Alzheimer's \& Dementia}, 20\penalty0 (5):\penalty0 3708--3821, 2024.
\newblock \doi{https://doi.org/10.1002/alz.13809}.
\newblock URL \url{https://alz-journals.onlinelibrary.wiley.com/doi/abs/10.1002/alz.13809}.

\bibitem[Silva et~al.(2019)Silva, Loures, Alves, {de Souza}, Borges, and Carvalho]{silvaAlzheimersDiseaseRisk2019}
Marcos Vin{\'i}cius~Ferreira Silva, Cristina de Mello~Gomide Loures, Luan Carlos~Vieira Alves, Leonardo~Cruz {de Souza}, Karina Braga~Gomes Borges, and Maria das~Gra{\c c}as Carvalho.
\newblock Alzheimer's disease: Risk factors and potentially protective measures.
\newblock \emph{Journal of Biomedical Science}, 26\penalty0 (1):\penalty0 33, May 2019.
\newblock ISSN 1423-0127.
\newblock \doi{10.1186/s12929-019-0524-y}.

\bibitem[{Avelar-Pereira} et~al.(2023){Avelar-Pereira}, Belloy, O'Hara, and Hosseini]{avelar-pereiraDecodingHeterogeneityAlzheimers2023}
B{\'a}rbara {Avelar-Pereira}, Michael~E. Belloy, Ruth O'Hara, and S.~M.~Hadi Hosseini.
\newblock Decoding the heterogeneity of {{Alzheimer}}'s disease diagnosis and progression using multilayer networks.
\newblock \emph{Molecular Psychiatry}, 28\penalty0 (6):\penalty0 2423--2432, June 2023.
\newblock ISSN 1476-5578.
\newblock \doi{10.1038/s41380-022-01886-z}.

\bibitem[Wang et~al.(2023)Wang, Yao, and Allen]{wangSupervisedConvexClustering2023}
Minjie Wang, Tianyi Yao, and Genevera~I. Allen.
\newblock Supervised convex clustering.
\newblock \emph{Biometrics}, 79\penalty0 (4):\penalty0 3846--3858, 2023.
\newblock ISSN 1541-0420.
\newblock \doi{10.1111/biom.13860}.

\bibitem[Zhang et~al.(2024)Zhang, Chang, and Long]{zhangRobustKnowledgeguidedBiclustering2024}
Qiyiwen Zhang, Changgee Chang, and Qi~Long.
\newblock Robust knowledge-guided biclustering for multi-omics data.
\newblock \emph{Briefings in Bioinformatics}, 25\penalty0 (1):\penalty0 bbad446, January 2024.
\newblock ISSN 1477-4054.
\newblock \doi{10.1093/bib/bbad446}.

\bibitem[Elman et~al.(2024)Elman, Schork, Rangan, and Initiative]{elmanExploringGeneticHeterogeneity2024}
Jeremy~A. Elman, Nicholas~J. Schork, Aaditya~V. Rangan, and Alzheimer's Disease~Neuroimaging Initiative.
\newblock Exploring the genetic heterogeneity of {{Alzheimer}}'s disease: {{Evidence}} for genetic subtypes.
\newblock \emph{medRxiv}, page 2023.05.02.23289347, May 2024.
\newblock \doi{10.1101/2023.05.02.23289347}.

\bibitem[Hartigan(1972)]{hartiganDirectClusteringData1972}
J.~A. Hartigan.
\newblock Direct {{Clustering}} of a {{Data Matrix}}.
\newblock \emph{Journal of the American Statistical Association}, 67\penalty0 (337):\penalty0 123--129, March 1972.
\newblock ISSN 0162-1459.
\newblock \doi{10.1080/01621459.1972.10481214}.

\bibitem[Sill et~al.(2011)Sill, Kaiser, Benner, and {Kopp-Schneider}]{sillRobustBiclusteringSparse2011}
Martin Sill, Sebastian Kaiser, Axel Benner, and Annette {Kopp-Schneider}.
\newblock Robust biclustering by sparse singular value decomposition incorporating stability selection.
\newblock \emph{Bioinformatics}, 27\penalty0 (15):\penalty0 2089--2097, August 2011.
\newblock ISSN 1367-4803.
\newblock \doi{10.1093/bioinformatics/btr322}.

\bibitem[Chekouo and Murua(2015)]{chekouoPenalizedBiclusteringModel2015}
Thierry Chekouo and Alejandro Murua.
\newblock The penalized biclustering model and related algorithms.
\newblock \emph{Journal of Applied Statistics}, 42\penalty0 (6):\penalty0 1255--1277, June 2015.
\newblock ISSN 0266-4763.
\newblock \doi{10.1080/02664763.2014.999647}.

\bibitem[Chekouo et~al.(2015)Chekouo, Murua, and Raffelsberger]{chekouoGibbsplaidBiclusteringModel2015}
Thierry Chekouo, Alejandro Murua, and Wolfgang Raffelsberger.
\newblock The {{Gibbs-plaid}} biclustering model.
\newblock \emph{The Annals of Applied Statistics}, 9\penalty0 (3):\penalty0 1643--1670, September 2015.
\newblock ISSN 1932-6157, 1941-7330.
\newblock \doi{10.1214/15-AOAS854}.

\bibitem[Li et~al.(2018)Li, Chang, Kundu, and Long]{liBayesianGeneralizedBiclustering2018}
Ziyi Li, Changgee Chang, Suprateek Kundu, and Qi~Long.
\newblock Bayesian generalized biclustering analysis via adaptive structured shrinkage.
\newblock \emph{Biostatistics (Oxford, England)}, 21\penalty0 (3):\penalty0 610--624, December 2018.
\newblock ISSN 1465-4644.
\newblock \doi{10.1093/biostatistics/kxy081}.

\bibitem[Zhang et~al.(2022)Zhang, Wendt, Bowler, Hersh, and Safo]{zhangRobustIntegrativeBiclustering2022}
Weijie Zhang, Christine Wendt, Russel Bowler, Craig~P Hersh, and Sandra~E Safo.
\newblock Robust integrative biclustering for multi-view data.
\newblock \emph{Statistical methods in medical research}, 31\penalty0 (11):\penalty0 2201--2216, November 2022.
\newblock ISSN 0962-2802.
\newblock \doi{10.1177/09622802221122427}.

\bibitem[Castanho et~al.(2024)Castanho, Aidos, and Madeira]{castanhoBiclusteringDataAnalysis2024}
Eduardo~N Castanho, Helena Aidos, and Sara~C Madeira.
\newblock Biclustering data analysis: A comprehensive survey.
\newblock \emph{Briefings in Bioinformatics}, 25\penalty0 (4):\penalty0 bbae342, July 2024.
\newblock ISSN 1467-5463.
\newblock \doi{10.1093/bib/bbae342}.

\bibitem[Yang et~al.(2017)Yang, Shen, Yuan, Zhang, and Wei]{yangAnalysisBreastCancer2017}
Liying Yang, Yunyan Shen, Xiguo Yuan, Junying Zhang, and Jianhua Wei.
\newblock Analysis of breast cancer subtypes by {{AP-ISA}} biclustering.
\newblock \emph{BMC Bioinformatics}, 18\penalty0 (1):\penalty0 481, November 2017.
\newblock ISSN 1471-2105.
\newblock \doi{10.1186/s12859-017-1926-z}.

\bibitem[Sun et~al.(2014)Sun, Bi, and Kranzler]{sunMultiviewSingularValue2014}
Jiangwen Sun, Jinbo Bi, and Henry~R. Kranzler.
\newblock Multi-view singular value decomposition for disease subtyping and genetic associations.
\newblock \emph{BMC Genetics}, 15\penalty0 (1):\penalty0 73, June 2014.
\newblock ISSN 1471-2156.
\newblock \doi{10.1186/1471-2156-15-73}.

\bibitem[Kurlowicz and Wallace(1999)]{kurlowiczMiniMentalStateExamination1999}
Lenore Kurlowicz and Meredith Wallace.
\newblock The {{Mini-Mental State Examination}} ({{MMSE}}).
\newblock \emph{Journal of Gerontological Nursing}, 25\penalty0 (5):\penalty0 8--9, May 1999.
\newblock ISSN 0098-9134, 1938-243X.
\newblock \doi{10.3928/0098-9134-19990501-08}.

\bibitem[Miyamoto et~al.(2008)Miyamoto, Ichihashi, and Honda]{miyamotoBasicMethodsCMeansClustering2008}
Sadaaki Miyamoto, Hidetomo Ichihashi, and Katsuhiro Honda.
\newblock {{BasicMethods}} for c-{{Means Clustering}}.
\newblock In Sadaaki Miyamoto, Hidetomo Ichihashi, and Katsuhiro Honda, editors, \emph{Algorithms for {{Fuzzy Clustering}}: {{Methods}} in c-{{Means Clustering}} with {{Applications}}}, pages 9--42. Springer, Berlin, Heidelberg, 2008.
\newblock ISBN 978-3-540-78737-2.
\newblock \doi{10.1007/978-3-540-78737-2_2}.

\bibitem[Wang and {Carreira-Perpi{\~n}{\'a}n}(2014)]{wangLaplacianKmodesAlgorithm2014}
Weiran Wang and Miguel~{\'A} {Carreira-Perpi{\~n}{\'a}n}.
\newblock The {{Laplacian K-modes}} algorithm for clustering, June 2014.

\bibitem[Bubeck(2015)]{bubeckConvexOptimizationAlgorithms2015}
S{\'e}bastien Bubeck.
\newblock Convex {{Optimization}}: {{Algorithms}} and {{Complexity}}.
\newblock \emph{Found. Trends Mach. Learn.}, 8\penalty0 (3-4):\penalty0 231--357, November 2015.
\newblock ISSN 1935-8237.
\newblock \doi{10.1561/2200000050}.

\bibitem[Wang and {Carreira-Perpi{\~n}{\'a}n}(2013)]{wangProjectionProbabilitySimplex2013}
Weiran Wang and Miguel~{\'A} {Carreira-Perpi{\~n}{\'a}n}.
\newblock Projection onto the probability simplex: {{An}} efficient algorithm with a simple proof, and an application, September 2013.

\bibitem[Chen and Chen(2008)]{chenExtendedBayesianInformation2008}
Jiahua Chen and Zehua Chen.
\newblock Extended {{Bayesian}} information criteria for model selection with large model spaces.
\newblock \emph{Biometrika}, 95\penalty0 (3):\penalty0 759--771, September 2008.
\newblock ISSN 0006-3444.
\newblock \doi{10.1093/biomet/asn034}.

\bibitem[Sun et~al.(2013)Sun, Miao, and Yan]{sunNoiseResistantBiclusterRecognition2013}
Huan Sun, Gengxin Miao, and Xifeng Yan.
\newblock Noise-{{Resistant Bicluster Recognition}}.
\newblock In \emph{2013 {{IEEE}} 13th {{International Conference}} on {{Data Mining}}}, pages 707--716, December 2013.
\newblock \doi{10.1109/ICDM.2013.34}.

\bibitem[Zhang et~al.(2021)Zhang, Tian, Wang, Ma, Tan, and Yu]{zhangEpidemiologyAlzheimersDisease2021}
X.-X. Zhang, Y.~Tian, Z.-T. Wang, Y.-H. Ma, Lan Tan, and Jin-Tai Yu.
\newblock The {{Epidemiology}} of {{Alzheimer}}'s {{Disease Modifiable Risk Factors}} and {{Prevention}}.
\newblock \emph{The Journal of Prevention of Alzheimer's Disease}, 8\penalty0 (3):\penalty0 313--321, July 2021.
\newblock ISSN 2426-0266.
\newblock \doi{10.14283/jpad.2021.15}.

\bibitem[Liebisch et~al.(2020)Liebisch, Fahy, Aoki, Dennis, Durand, Ejsing, Fedorova, Feussner, Griffiths, K{\"o}feler, Merrill, Murphy, O'Donnell, Oskolkova, Subramaniam, Wakelam, and Spener]{liebischUpdateLIPIDMAPS2020}
Gerhard Liebisch, Eoin Fahy, Junken Aoki, Edward~A. Dennis, Thierry Durand, Christer~S. Ejsing, Maria Fedorova, Ivo Feussner, William~J. Griffiths, Harald K{\"o}feler, Alfred~H. Merrill, Robert~C. Murphy, Valerie~B. O'Donnell, Olga Oskolkova, Shankar Subramaniam, Michael J.~O. Wakelam, and Friedrich Spener.
\newblock Update on {{LIPID MAPS}} classification, nomenclature, and shorthand notation for {{MS-derived}} lipid structures.
\newblock \emph{Journal of Lipid Research}, 61\penalty0 (12):\penalty0 1539--1555, December 2020.
\newblock ISSN 0022-2275, 1539-7262.
\newblock \doi{10.1194/jlr.S120001025}.

\bibitem[Devanand et~al.(2012)Devanand, Bansal, Liu, Hao, Pradhaban, and Peterson]{devanandMRIHippocampalEntorhinal2012}
D.~P. Devanand, Ravi Bansal, Jun Liu, Xuejun Hao, Gnanavalli Pradhaban, and Bradley~S. Peterson.
\newblock {{MRI}} hippocampal and entorhinal cortex mapping in predicting conversion to {{Alzheimer}}'s disease.
\newblock \emph{Neuroimage}, 60\penalty0 (3):\penalty0 1622--1629, April 2012.
\newblock ISSN 1053-8119.
\newblock \doi{10.1016/j.neuroimage.2012.01.075}.

\bibitem[Cai et~al.(2015)Cai, Chong, Zhang, Li, {von Deneen}, Ren, Dong, and Huang]{caiAlteredFunctionalConnectivity2015}
Suping Cai, Tao Chong, Yun Zhang, Jun Li, Karen~M. {von Deneen}, Junchan Ren, Minghao Dong, and Liyu Huang.
\newblock Altered {{Functional Connectivity}} of {{Fusiform Gyrus}} in {{Subjects}} with {{Amnestic Mild Cognitive Impairment}}: {{A Resting-State fMRI Study}}.
\newblock \emph{Frontiers in Human Neuroscience}, 9:\penalty0 471, August 2015.
\newblock ISSN 1662-5161.
\newblock \doi{10.3389/fnhum.2015.00471}.

\bibitem[Guan et~al.(2017)Guan, Liu, Jiang, Tao, Zhang, Niu, Zhu, Wang, Cheng, Kochan, Brodaty, Sachdev, and Wen]{guanClassifyingMCISubtypes2017}
Hao Guan, Tao Liu, Jiyang Jiang, Dacheng Tao, Jicong Zhang, Haijun Niu, Wanlin Zhu, Yilong Wang, Jian Cheng, Nicole~A. Kochan, Henry Brodaty, Perminder Sachdev, and Wei Wen.
\newblock Classifying {{MCI Subtypes}} in {{Community-Dwelling Elderly Using Cross-Sectional}} and {{Longitudinal MRI-Based Biomarkers}}.
\newblock \emph{Frontiers in Aging Neuroscience}, 9:\penalty0 309, September 2017.
\newblock ISSN 1663-4365.
\newblock \doi{10.3389/fnagi.2017.00309}.

\bibitem[{van de Mortel} et~al.(2021){van de Mortel}, Thomas, and {van Wingen}]{vandemortelGreyMatterLoss2021}
Laurens~Ansem {van de Mortel}, Rajat~Mani Thomas, and Guido~Alexander {van Wingen}.
\newblock Grey {{Matter Loss}} at {{Different Stages}} of {{Cognitive Decline}}: {{A Role}} for the {{Thalamus}} in {{Developing Alzheimer}}'s {{Disease}}.
\newblock \emph{Journal of Alzheimer's Disease}, 83\penalty0 (2):\penalty0 705--720, September 2021.
\newblock ISSN 1387-2877.
\newblock \doi{10.3233/JAD-210173}.

\bibitem[Sachdev et~al.(2013)Sachdev, Zhuang, Braidy, and Wen]{sachdevAlzheimersDiseaseWhite2013}
Perminder~S. Sachdev, Lin Zhuang, Nady Braidy, and Wei Wen.
\newblock Is {{Alzheimer}}'s a disease of the white matter?
\newblock \emph{Current Opinion in Psychiatry}, 26\penalty0 (3):\penalty0 244, May 2013.
\newblock ISSN 0951-7367.
\newblock \doi{10.1097/YCO.0b013e32835ed6e8}.

\bibitem[Akyol et~al.(2021)Akyol, Ugur, Yilmaz, Ustun, Gorti, Oh, McGuinness, Passmore, Kehoe, Maddens, Green, and Graham]{akyolLipidProfilingAlzheimers2021}
Sumeyya Akyol, Zafer Ugur, Ali Yilmaz, Ilyas Ustun, Santosh Kapil~Kumar Gorti, Kyungjoon Oh, Bernadette McGuinness, Peter Passmore, Patrick~G. Kehoe, Michael~E. Maddens, Brian~D. Green, and Stewart~F. Graham.
\newblock Lipid {{Profiling}} of {{Alzheimer}}'s {{Disease Brain Highlights Enrichment}} in {{Glycerol}}(phospho)lipid, and {{Sphingolipid Metabolism}}.
\newblock \emph{Cells}, 10\penalty0 (10):\penalty0 2591, September 2021.
\newblock ISSN 2073-4409.
\newblock \doi{10.3390/cells10102591}.

\bibitem[Wood(2012)]{woodLipidomicsAlzheimersDisease2012}
Paul~L Wood.
\newblock Lipidomics of {{Alzheimer}}'s disease: Current status.
\newblock \emph{Alzheimer's Research \& Therapy}, 4\penalty0 (1):\penalty0 5, February 2012.
\newblock ISSN 1758-9193.
\newblock \doi{10.1186/alzrt103}.

\bibitem[Lock et~al.(2013)Lock, Hoadley, Marron, and Nobel]{lockJointIndividualVariation2013}
Eric~F. Lock, Katherine~A. Hoadley, J.~S. Marron, and Andrew~B. Nobel.
\newblock Joint and individual variation explained ({{JIVE}}) for integrated analysis of multiple data types.
\newblock \emph{The Annals of Applied Statistics}, 7\penalty0 (1):\penalty0 523--542, March 2013.
\newblock ISSN 1932-6157, 1941-7330.
\newblock \doi{10.1214/12-AOAS597}.

\end{thebibliography}


\begin{thebibliography}{9}
\providecommand{\natexlab}[1]{#1}
\providecommand{\url}[1]{\texttt{#1}}
\expandafter\ifx\csname urlstyle\endcsname\relax
  \providecommand{\doi}[1]{doi: #1}\else
  \providecommand{\doi}{doi: \begingroup \urlstyle{rm}\Url}\fi

\bibitem[Li et~al.(2018)Li, Chang, Kundu, and Long]{liBayesianGeneralizedBiclustering2018}
Ziyi Li, Changgee Chang, Suprateek Kundu, and Qi~Long.
\newblock Bayesian generalized biclustering analysis via adaptive structured shrinkage.
\newblock \emph{Biostatistics (Oxford, England)}, 21\penalty0 (3):\penalty0 610--624, December 2018.
\newblock ISSN 1465-4644.
\newblock \doi{10.1093/biostatistics/kxy081}.

\bibitem[Wang and {Carreira-Perpi{\~n}{\'a}n}(2013)]{wangProjectionProbabilitySimplex2013}
Weiran Wang and Miguel~{\'A} {Carreira-Perpi{\~n}{\'a}n}.
\newblock Projection onto the probability simplex: {{An}} efficient algorithm with a simple proof, and an application, September 2013.

\bibitem[Chen and Chen(2008)]{chenExtendedBayesianInformation2008}
Jiahua Chen and Zehua Chen.
\newblock Extended {{Bayesian}} information criteria for model selection with large model spaces.
\newblock \emph{Biometrika}, 95\penalty0 (3):\penalty0 759--771, September 2008.
\newblock ISSN 0006-3444.
\newblock \doi{10.1093/biomet/asn034}.

\bibitem[Chen and Chen(2012)]{chenExtendedBICSmallnlargeP2012}
Jiahua Chen and Zehua Chen.
\newblock Extended {{BIC}} for small-n-large-{{P}} sparse {{GLM}}.
\newblock \emph{Statistica Sinica}, 22\penalty0 (2), April 2012.
\newblock ISSN 10170405.
\newblock \doi{10.5705/ss.2010.216}.

\bibitem[Zhang et~al.(2022)Zhang, Wendt, Bowler, Hersh, and Safo]{zhangRobustIntegrativeBiclustering2022}
Weijie Zhang, Christine Wendt, Russel Bowler, Craig~P Hersh, and Sandra~E Safo.
\newblock Robust integrative biclustering for multi-view data.
\newblock \emph{Statistical methods in medical research}, 31\penalty0 (11):\penalty0 2201--2216, November 2022.
\newblock ISSN 0962-2802.
\newblock \doi{10.1177/09622802221122427}.

\bibitem[Lee et~al.(2010)Lee, Shen, Huang, and Marron]{leeBiclusteringSparseSingular2010}
Mihee Lee, Haipeng Shen, Jianhua~Z. Huang, and J.~S. Marron.
\newblock Biclustering via {{Sparse Singular Value Decomposition}}.
\newblock \emph{Biometrics}, 66\penalty0 (4):\penalty0 1087--1095, 2010.
\newblock ISSN 1541-0420.
\newblock \doi{10.1111/j.1541-0420.2010.01392.x}.

\bibitem[Sill et~al.(2011)Sill, Kaiser, Benner, and {Kopp-Schneider}]{sillRobustBiclusteringSparse2011}
Martin Sill, Sebastian Kaiser, Axel Benner, and Annette {Kopp-Schneider}.
\newblock Robust biclustering by sparse singular value decomposition incorporating stability selection.
\newblock \emph{Bioinformatics}, 27\penalty0 (15):\penalty0 2089--2097, August 2011.
\newblock ISSN 1367-4803.
\newblock \doi{10.1093/bioinformatics/btr322}.

\bibitem[Meinshausen and B{\"u}hlmann(2010)]{meinshausenStabilitySelection2010}
Nicolai Meinshausen and Peter B{\"u}hlmann.
\newblock Stability {{Selection}}.
\newblock \emph{Journal of the Royal Statistical Society Series B: Statistical Methodology}, 72\penalty0 (4):\penalty0 417--473, September 2010.
\newblock ISSN 1369-7412.
\newblock \doi{10.1111/j.1467-9868.2010.00740.x}.

\bibitem[Sun et~al.(2013)Sun, Miao, and Yan]{sunNoiseResistantBiclusterRecognition2013}
Huan Sun, Gengxin Miao, and Xifeng Yan.
\newblock Noise-{{Resistant Bicluster Recognition}}.
\newblock In \emph{2013 {{IEEE}} 13th {{International Conference}} on {{Data Mining}}}, pages 707--716, December 2013.
\newblock \doi{10.1109/ICDM.2013.34}.

\end{thebibliography}






\section*{Supporting Information}

We present additional information referenced in Section~\ref{s:model}, Section~\ref{s:sim}, and Section~\ref{s:rda} in the Supporting Information. \vspace*{-8pt}






\end{document}


\title{Supporting Information for ``Supervised Integrative Biclustering with applications to Alzheimer’s Disease''}
\date{\vspace{-10ex}}
\maketitle

\section{More on Methods}

\subsection{More on Exponential Family}
\par We summarize the mean, natural parameter, and choice of function $G$ for common distributions from the exponential family in Web Table A.1.
\begin{table}[h]
\label{tab:natpar}
\begin{center}
\begin{tabular}{|c|c|c|c|}
    \hline
    Distribution & Mean & Natural parameter $\psi$ & cumulant function $G(\psi)$ \\
    \hline
    Gaussian & $\mu$ & $\mu$ & $\psi^2/2$\\
    \hline
    Bernoulli & $p$ & $\log(\frac{p}{1-p})$ & $\log(1+\exp(\psi))$ \\
    \hline
    Poisson & $\lambda$ & $\log(\lambda)$ & $\exp(\psi)$\\
    \hline
\end{tabular}
\end{center}
\caption{Mean, natural parameters, and convex cumulant functions of commonly used distributions from the exponential family distribution}
\end{table}

\subsection{More on Parameter Estimation}

\par The parameters are estimated such that $\bm{U,V,W,\mu,\beta}=\text{argmin }l(\bm{X,y,U,V,W,\mu,\beta})$. For the initialization part, we followed ideas in GBC  \citep{liBayesianGeneralizedBiclustering2018} and initialized $\bm{\Psi^{(d)}}$ according to their corresponding distribution using the following function:

\begin{equation*}
    \bm{\Psi_{ij}^{(d)}}= 
    \begin{cases}
        X_{ij}^{(d)},& \text{if } X \text{ is from a normal distribution}\\
        \text{logit}(\frac{X_{ij}^{(d)}+1}{3}),& \text{if } X \text{ is from a Bernoulli distribution}\\
        \text{logit}(\frac{X_{ij}^{(d)}+1}{r_j+X_{ij}^{(d)}+2}),& \text{if } X \text{ is from a Negative Binomial distribution}\\
        \text{log}(X_{ij}^{(d)}+1),& \text{if } X \text{ is from a Poisson distribution},
    \end{cases}
\end{equation*}

where $r_j$ is the number of failures for the negative binomial distribution.

\par When updating $\bm{V}^{(d)}$, we use the proximal map for $L_1$ penalty, the soft-thresholding function, after applying the gradient descent step. The soft-thresholding function $\mathcal{S}(z,\lambda)$ is defined as 

\begin{equation*}
    \mathcal{S}(z,\lambda)= 
    \begin{cases}
        z-\lambda,& \text{if } z\geq\lambda\\
        0,& -\lambda\leq z<\lambda\\
        z+\lambda,& \text{if } z<-\lambda \\
    \end{cases}
\end{equation*}

\par We project each row of $\bm{W}$ onto a probability simplex using the algorithm proposed by \cite{wangProjectionProbabilitySimplex2013}. Section D in the supporting information provides algorithms to estimate these parameters and make predictions.

\subsection{Hyper-parameters Selection}
\par In this subsection, we describe our approach to search for the optimal set of hyper-parameters $\lambda_{kd}$'s and $K$, the number of biclusters. We propose to use a variant of the Bayesian information criterion (BIC) to select the best set of hyper-parameters $\lambda_{kd}$'s. This variant of BIC is defined as $\text{BIC} = q\log(n) - 2 \log(\widehat{L})$, where $q$ is the total number of variables selected in all views, $n$ is the sample size of the data, and $\widehat{L}$ is the total loss with the estimated parameters. Note that this $\widehat{L}$ will be different from the loss function we estimated because our loss function is based on an average, and we provide options for users to decide whether they would like to use the differentiable part of the original loss function $\widehat{l_0}$ or the total loss $\widehat{L}$. We use random search to facilitate the process of tuning, and the set of tuning parameters with the lowest value of BIC are selected. By default we search for the maximum of the search space size and $60$ time so that the probability of choosing a result from the top $5\%$ percentile within the search space is greater than 0.95.

\par Besides this variant of BIC, we also provide a variant of BIC, the extended BIC (eBIC), to give flexibility for the users \citep{chenExtendedBayesianInformation2008}. Since BIC is in favor of results with better loss function values but low sparsity, especially when the dimension is high, we introduce eBIC defined as $\text{eBIC} =- 2 \log(\widehat{L}) + q\log(n) + 2\sigma \sum_{d=1}^D q\log(p^{(d)})$, to address this problem. The additional term $2\sigma q\log(p^{(d)})$ accounts for the number of plausible combinations of the same size. $0\leq\sigma\leq1$ can be chosen to further decide the extent of sparsity, and when $\sigma=0$ this becomes a regular BIC. The original version requires calculating permutation and combinations, which can be computationally expensive, so we choose to use this simplified version proposed in a later paper from the same author to simplify the computation \citep{chenExtendedBICSmallnlargeP2012}.

\par We use the number of biclusters specified by the user if it is supplied. Otherwise, we search for it in a range from small to large until there is an empty bicluster that contains no sample or variable. Since the number of biclusters is given by the number of columns of the matrix $\bm{W}$ (or equivalently the number of columns of the matrix $\bm{V}^{(d)}$), it is possible to have several columns full of zeros in $\bm{W}$ and $\bm{V}^{(d)}$. In that case, we terminate the search and conclude that the number of biclusters is given by the largest one in the search process that does not result in any empty bicluster.

\section{More on Simulation}
\subsection{Details on Simulation Settings}
\par For all of the simulation settings, the data in all views are generated as follows: We first simulate the mean from the Gaussian distribution $\bm{\Psi}^{(d)}$ for each view $d$ by $\bm{\Psi}^{(d)} = \bm{1\mu}^{(d)T}+(\bm{U}\cdot \bm{W})\bm{S}(\bm{V}^{(d)}\cdot \bm{\Gamma}^{(d)T})$, where $\bm{U}, \bm{W}\in \mathbb{R}^{n\times K}$, $\bm{V}^{(d)},\bm{\Gamma}^{(d)}\in \mathbb{R}^{p^{(d)}\times K}$, $\bm{S}\in \mathbb{R}^{K\times K}$. Each entry of $\bm{\mu}^{(d)}$ follows a standard normal distribution. Each entry of $\bm{U}$ and $\bm{V}$ is from a uniform distribution $U(0.5,1)$. $\bm{W}$ is simulated so that in each row, there is exactly one entry with value one, and the other entries are zeros. The entry with value one represents the bicluster that it belongs to. $\bm{\Gamma}$ controls the variable clusters. Since we assume the variable cluster to be sparse and the underlying column structure to be nonoverlapping, we simulate $\bm{\Gamma}$ such that each row can have at most one entry with value one, and each column have exactly $l$ entries of one, indicating important variables. We fix the number of true biclusters to be $K=3$ and vary the number of samples and the number of variables in each view. We further assume that the number of variables in each view is the same for all views. We use the following three combinations, $(n=150, p^{(d)}=100)$, $(n=150, p^{(d)}=500)$, and $(n=500, p^{(d)}=1000)$, to show the ability to recover the underlying structure of our methods under different sample sizes and different ratio of variables to samples. Motivated by \cite{zhangRobustIntegrativeBiclustering2022}, we introduce a term $\bm{S}$ that acts like singular values and fix it as $\bm{S}=\text{diag}(27,15,10)$ for all simulation settings to make the biclusters detectable. $10\%$ of the variables in each view are assumed to be important variables for each bicluster. For example, when $p^{(d)}=100$, we set $l=10$ for each bicluster in each view, and the total number of important variables in each view is $Kl=30$. Lastly, we simulate $x_{ij}^{(d)}\sim N(\psi_{ij}^{(d)}, \sigma^2)$, where we fix $\sigma^2=1$.

\par To simulate the outcome $\bm{y}$, we use similar steps for generating the $\bm{X}$. We first simulate the natural parameter $\bm{\psi_y = W\beta}$, where $\bm{W}$ is already obtained when simulating $\bm{X}$, and we fix $\bm{\beta} = (1,-1,-5)$ when $y$ is continuous. When $y$ is binary, we fix $\bm{\beta} = (1.5,0,-1.5)$ to ensure that the number of positive and negative samples is reasonable in all biclusters, especially when the sample size is small. After obtaining the natural parameter, we plug them into different distributions according to the simulation settings. For a continuous outcome, we let each entry $y_i$ in $\bm{y}$ be simulated from $N(\psi_{y_i}, \sigma_y^2)$, where $\sigma_y^2=1$ is fixed. For a binary outcome, we simulate $y_i\sim \text{Ber}(\frac{e^{\psi_{y_i}}}{1+e^{\psi_{y_i}}})$. Web Figure B.1 shows an example of the simulated multi-view data on a heatmap and simulated outcomes shown by a boxplot and a barplot.

\begin{figure}[!ht]
  \centering
  \subfloat[][]{\includegraphics[width=\textwidth]{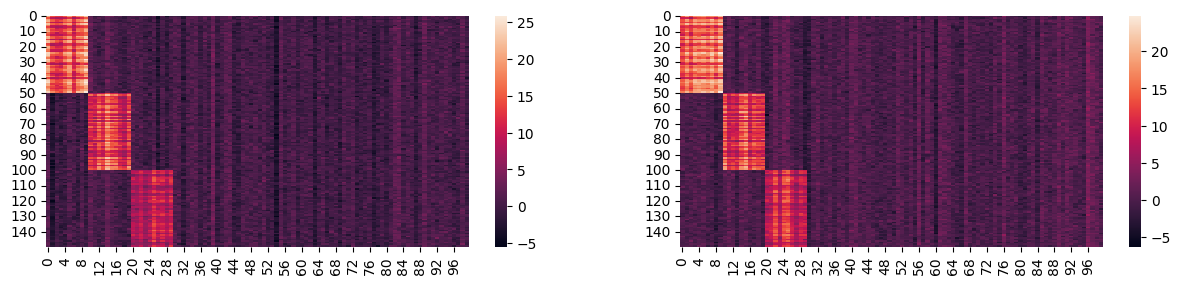}}\\
  \subfloat[][]{\includegraphics[width=.4\textwidth]{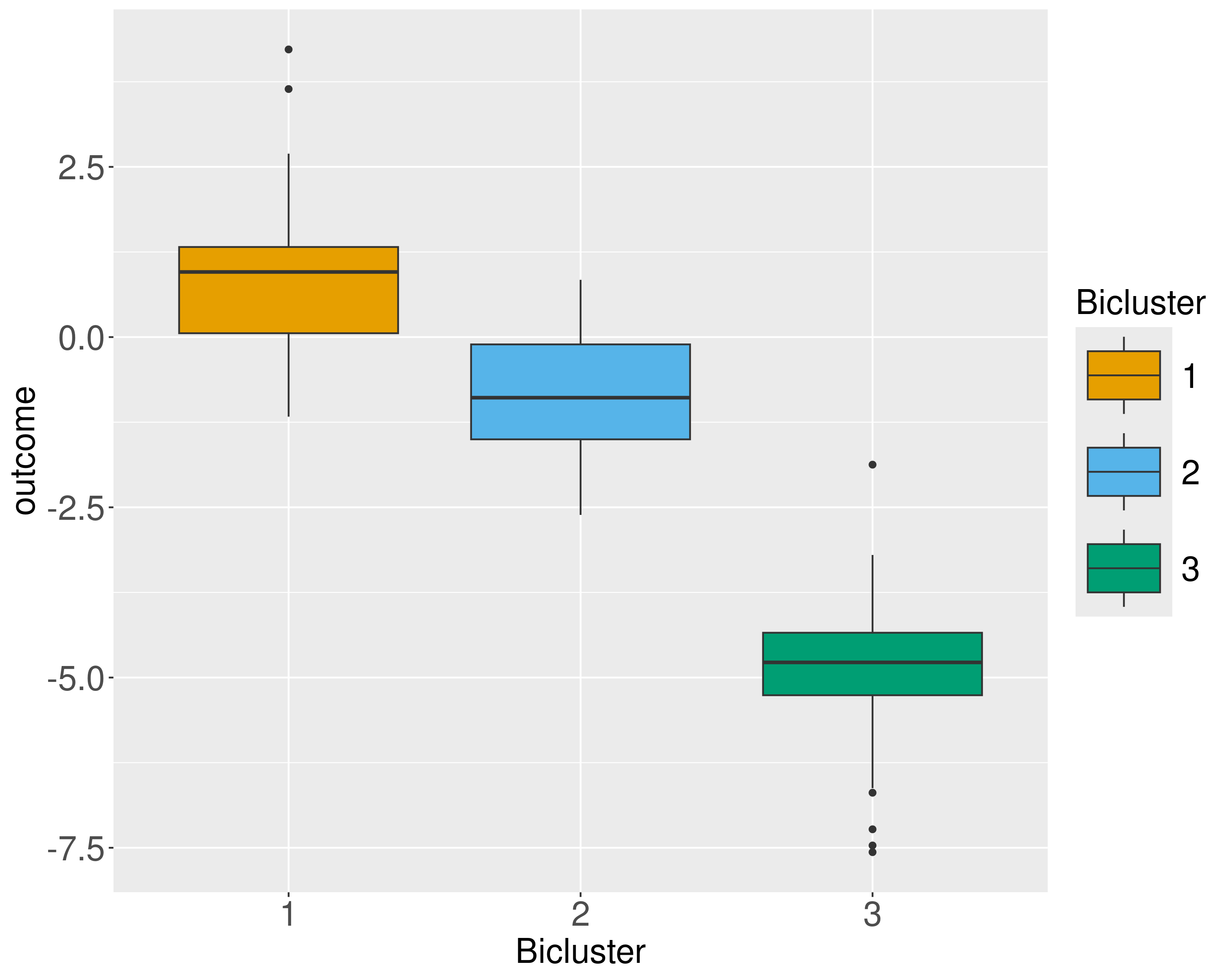}}\quad
  \subfloat[][]{\includegraphics[width=.4\textwidth]{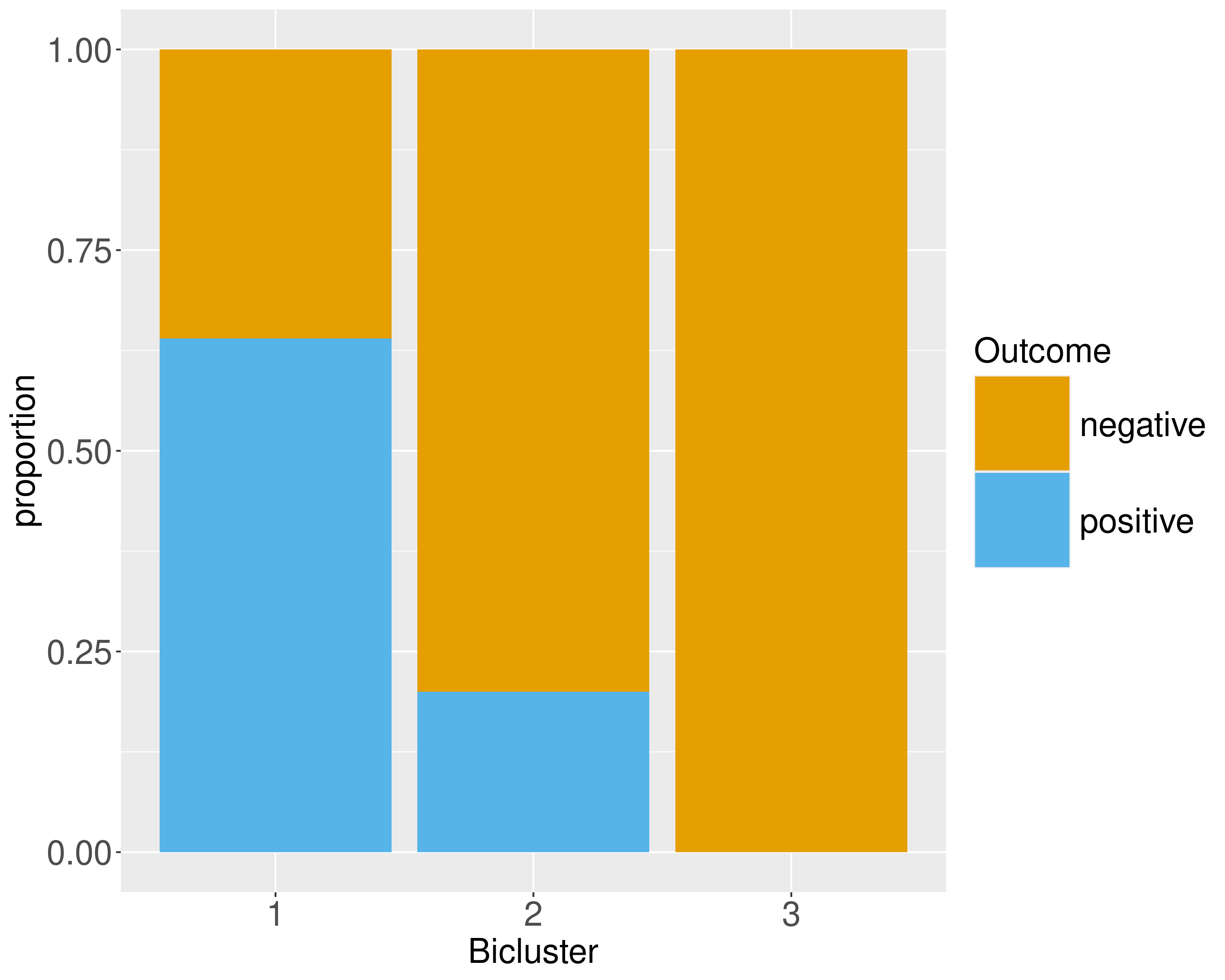}
  }
  \caption{The figures are plotted using one of the repetitions with 150 samples and 100 variables in each view. Heatmaps in (a) show the values of data in two views. The heatmaps are sorted using the true bicluster structures for better clarity. The boxplot in (b) and the bar plot in (c) show the distribution of the continuous outcome and the binary outcome respectively.}
  \label{fig:sim_example}
\end{figure}

\par Furthermore, we generate an independent testing set of the same size as the training data to assess the performance of predicting the outcome on new samples. The testing data is generated using the same $\bm{V}^{(d)}, \bm{\Gamma}^{(d)T}$, and $\bm{\mu}^{(d)}$, representing the same underlying variable structures as the simulated training data. A different pair of $\bm{U}$ and $\bm{W}$ is generated under the same distribution as the training data, indicating the bicluster that the new samples belong to. We keep the $\bm{\beta}$ to be the same as the training outcome and generate testing outcomes using the new $\bm{W}$. We use the prediction method mentioned in Section 2.5 to predict which bicluster the new samples belong to and the outcome $\bm{y}$ for our proposed method.

\subsection{Brief Introduction of Comparison Methods}
\par GBC achieves multiview biclustering by concatenating data from all the views together and specifying the data type of each column of the data \citep{liBayesianGeneralizedBiclustering2018}. It achieves biclustering by computing a latent factor loading matrix $\bm{W}$ and a latent factor matrix $\bm{Z}$ that are both sparse. The bicluster is defined by the non-zero entries of $\bm{W}$ and $\bm{Z}$: the non-zero elements in $\bm{Z}_{\cdot k}$ represent the subset of subjects belonging to the $k$th bicluster, and non-zero elements in $\bm{W}_{\cdot k}$ represent the subset of features that contribute to the $k$th bicluster. GBC only allows overlapping biclustering structure because of its formulation, and the number of biclusters must be specified prior to computation. In our comparison, we specify the correct number of biclusters for GBC and used BIC to select the optimal set of tuning parameter as suggested in the original paper.

\par SSVD obtains sparse pairs of singular vectors that can be interpreted as biclusters by using sparsity-inducing penalties to a regular SVD \citep{leeBiclusteringSparseSingular2010}. The first pair of singular vectors $u_1,v_1$ and the scalar $d_1$ best approximates the original data matrix $\bm{X}$ with the inclusion of penalty terms, and the $i^{\text{th}}$ pairs of singular vectors approximate the remainder of $\bm{X}-\sum_{k=1}^{i-1}d_ku_kv_k^{T}$ with the penalty terms. Then, the non-zero entries in $u_i$ and $v_i$ naturally clustered rows and columns together, respectively, and these rows and columns form the $i^{\text{th}}$ bicluster. However, SSVD method does not give a stopping criterion for finding the sparse SVD layers, so the number of biclusters is arbitrary. S4VD extends SVD by adding stability selection to control Type I error rates in selecting variables and to provide a stopping criterion for detecting the number of biclusters \citep{sillRobustBiclusteringSparse2011}. Stability selection combines resampling with variable selection to choose the penalization parameters and to control the Type I error rate of the SVD layers \citep{meinshausenStabilitySelection2010}. It estimates the probability of being selected for all the variables and chooses the variables whose probabilities are greater than an a priori-specified threshold for each set of penalization parameters. The S4VD algorithm stops when there is no more stable variable selected, and this determines the number of biclusters identified. iSSVD further extends S4VD by taking into account the rich information from data in multiple views \citep{zhangRobustIntegrativeBiclustering2022}. This allows the results to be more interpretable than simply stacking the datasets. In addition, it uses all the data, instead of just the unclustered samples in SSVD and S4VD, when estimating the subsequent biclusters. The tuning parameters for both methods were computed using the formula given in these methods.

\subsection{Evaluation Methods}
\par Suppose $\bm{\bm{M}}^{(d)}=\{\bm{M}_1^{(d)},\dots, \bm{M}_K^{(d)}\}$ and $\bm{M}^{(d)*}=\{\bm{M}^{(d)*}_1,\dots, \bm{M}^{(d)*}_{K^*}\}$ are the sets of estimated and true biclusters for the $d^\text{th}$ view, respectively. Each of the $\bm{M}^{(d)}_k$ or $\bm{M}^{(d)*}_k$ is a $n\times p^{(d)}$ matrix where the entries ${\bm{M}_t}(i,j)=1$ if the $i^\text{th}$ sample and the $j^\text{th}$ variable is in the $t^\text{th}$ bicluster and $\bm{M}_{t}(i,j)=0$ otherwise. The Jaccard index for two biclusters is defined as 
\begin{align*}
    Jac(\bm{M}^{(d)}_k,\bm{M}^{(d)*}_{k^*})=\frac{|\bm{M}^{(d)}_k\cap \bm{M}^{(d)*}_{k^*}|}{|\bm{M}^{(d)}_k\cup \bm{M}^{(d)*}_{k^*}|},
\end{align*}
where $|\bm{M}^{(d)}_k\cap \bm{M}^{(d)*}_{k^*}|$ is the number of entries that are one in both $\bm{M}^{(d)}_k$ and $\bm{M}^{(d)*}_{k^*}$, and $|\bm{M}^{(d)}_k\cup \bm{M}^{(d)*}_{k^*}|$ is the number of entries that are one in at least one of $\bm{M}^{(d)}_k$ and $\bm{M}^{(d)*}_{k^*}$. We report the average relevance, recovery scores, and F-score using the definitions commonly used in other biclustering methods  \citep{sillRobustBiclusteringSparse2011, zhangRobustIntegrativeBiclustering2022, sunNoiseResistantBiclusterRecognition2013}: 
\begin{align*}
    \text{Relevance} &= \frac{1}{D}\sum_{d=1}^D\frac{1}{K}\sum_{k=1}^{K}\max_{k^*\in \{1,2,\dots,K^*\}} Jac(\bm{M}^{(d)}_k,\bm{M}^{(d)*}_{k^*})\\
    \text{Recovery} &= \frac{1}{D}\sum_{d=1}^D\frac{1}{K^*}\sum_{k^*=1}^{K^*}\max_{k\in \{1,2,\dots,K\}} Jac(\bm{M}^{(d)}_k,\bm{M}^{(d)*}_{k^*})\\
    \text{F-score} &= \frac{2\times \text{Relevance} \times \text{Recovery}}{\text{Relevance} + \text{Recovery}}
\end{align*}

\par False Negatives and False Positives in biclusters are also reported to ensure the comprehension of our evaluation. They are defined using $\bm{M}^{(d)}$ and $\bm{M}^{(d)*}$. Within each view $d$, for each pair of true bicluster and estimated bicluster $k$ and $k^*$, we record a false positive and false negative matrix $\bm{FP}^{(d)}_{k,k^*}$ and $\bm{FN}^{(d)}_{k,k^*}$. The false positive matrix for each pair is a $n\times p^{(d)}$ matrix where the $(i,j)^{\text{th}}$ entry of $\bm{FP}^{(d)}_{k,k^*}(i,j) = 1$ if the $(i,j)^{\text{th}}$ entry of the true bicluster $\bm{M}^{(d)}_t1(i,j)=0$ and the $(i,j)^{\text{th}}$ entry of the estimated bicluster $\bm{M}^{(d)*}_t2(i,j)=1$, and $\bm{FP}^{(d)}_{k,k^*}(i,j) = 0$ otherwise. Similarly, the false negative matrix is a $n\times p^{(d)}$ matrix where $FN^{(d)}_{k,k^*}(i,j) = 1$ if $\bm{M}^{(d)}_t1(i,j)=1$ and $\bm{M}^{(d)*}_t2(i,j)=0$, and $\bm{FN}^{(d)}_{k,k^*}(i,j) = 0$ otherwise. Then, for each view, we can construct a false positive matrix $\bm{FP}^{(d)}$ and a false negative matrix $\bm{FN}^{(d)}$. The false positive matrix for $d^{\text{th}}$ view $\bm{FP}^{(d)}$ is a $K\times K^*$ matrix where $\bm{FP}^{(d)}(i,j)$ is the mean of all the entries in $\bm{FP}^{(d)}_{k,k^*}$; similarly, the false negative matrix for $d^{\text{th}}$ view $\bm{FN}^{(d)}$ is a $K\times K^*$ matrix where $\bm{FN}^{(d)}(i,j)$ is the mean of all the entries in $\bm{FN}^{(d)}_{k,k^*}$. We calculate false positives and false negatives based on the true biclusters. The minimum in each row of $\bm{FP}^{(d)}$ and $\bm{FN}^{(d)}$ are recorded as the false positive and false negative for each bicluster, respectively. The mean of all the false positives and false negatives in all views are reported as the false positives and false negatives.

\subsection{Additional Results}

\par We include additional results from our simulation in this subsection. Web Figure B.1-B.4 summarizes the training and testing performances for all the methods when the outcome is continuous and binary, respecitvely. In Web Table 2, we also compare the average computation time required for all the methods we considered in the simulation.

\begin{figure}[H]
\includegraphics[width=\textwidth]{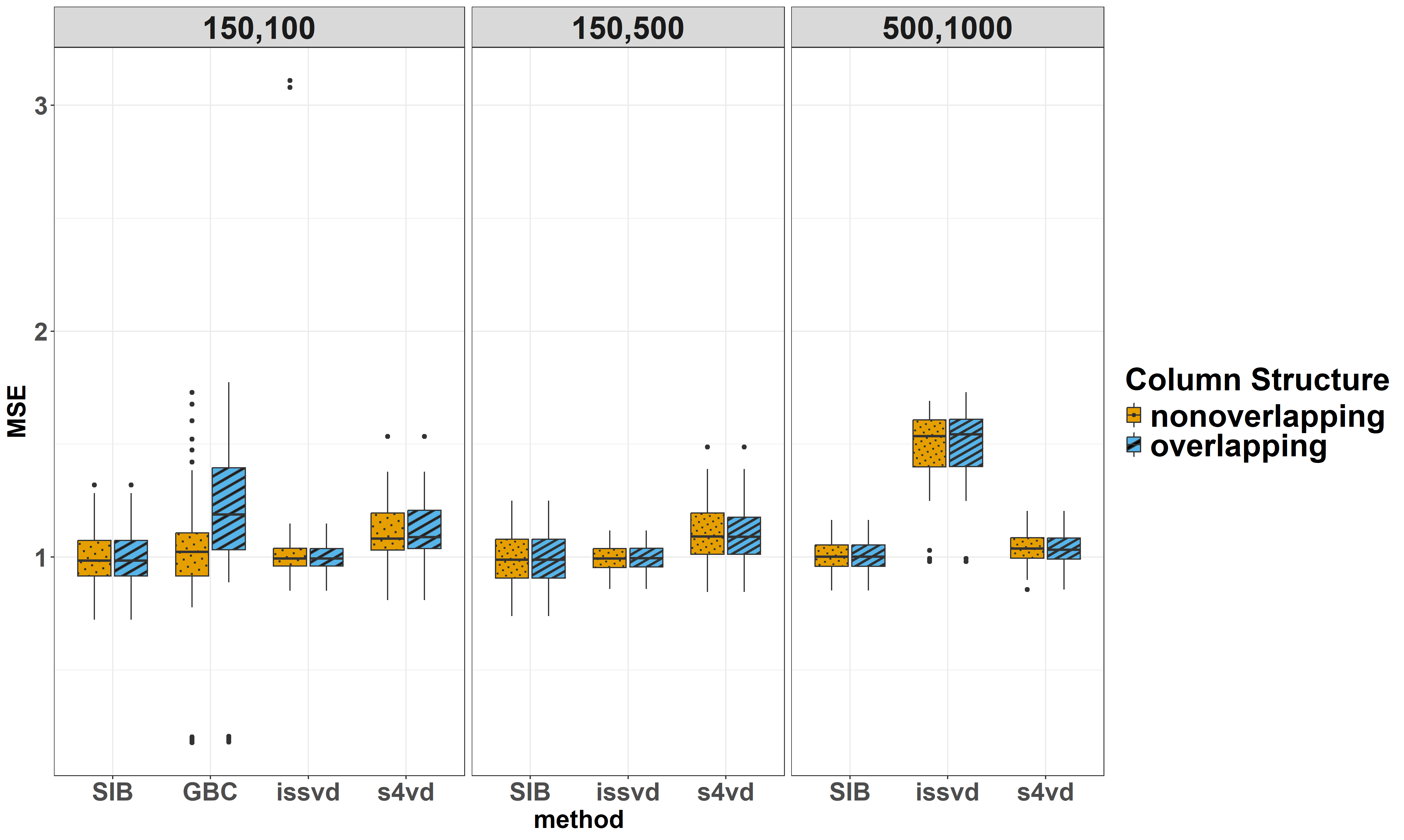}
\caption{Boxplot of the training mean square error (MSE) for all methods. The title indicates the number of samples and variables per view. The color of the boxes indicates whether or not we assume the columns to be nonoverlapping when estimating the biclusters.}
\label{fig:box_gg_MSE}
\end{figure}

\begin{figure}[H]
\includegraphics[width=\textwidth]{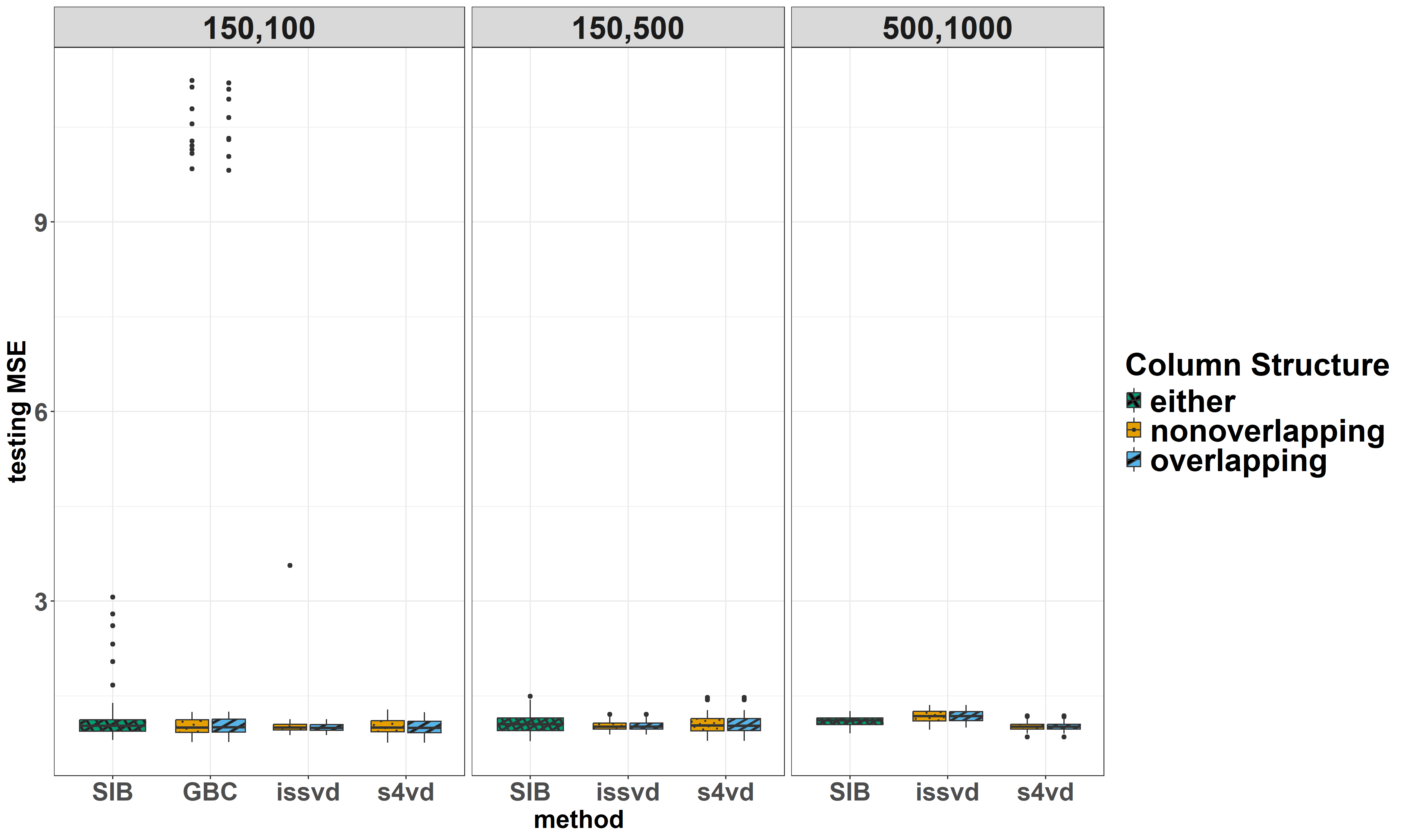}
\caption{Boxplot of the testing mean square error (MSE) for all methods. The title indicates the number of samples and variables per view. The color of the boxes indicates whether or not we assume the columns to be nonoverlapping when estimating the biclusters. Since the prediction with our proposed method does not depend on column structures, it is labeled as "either".}
\end{figure}

\begin{figure}[H]
\includegraphics[width=\textwidth]{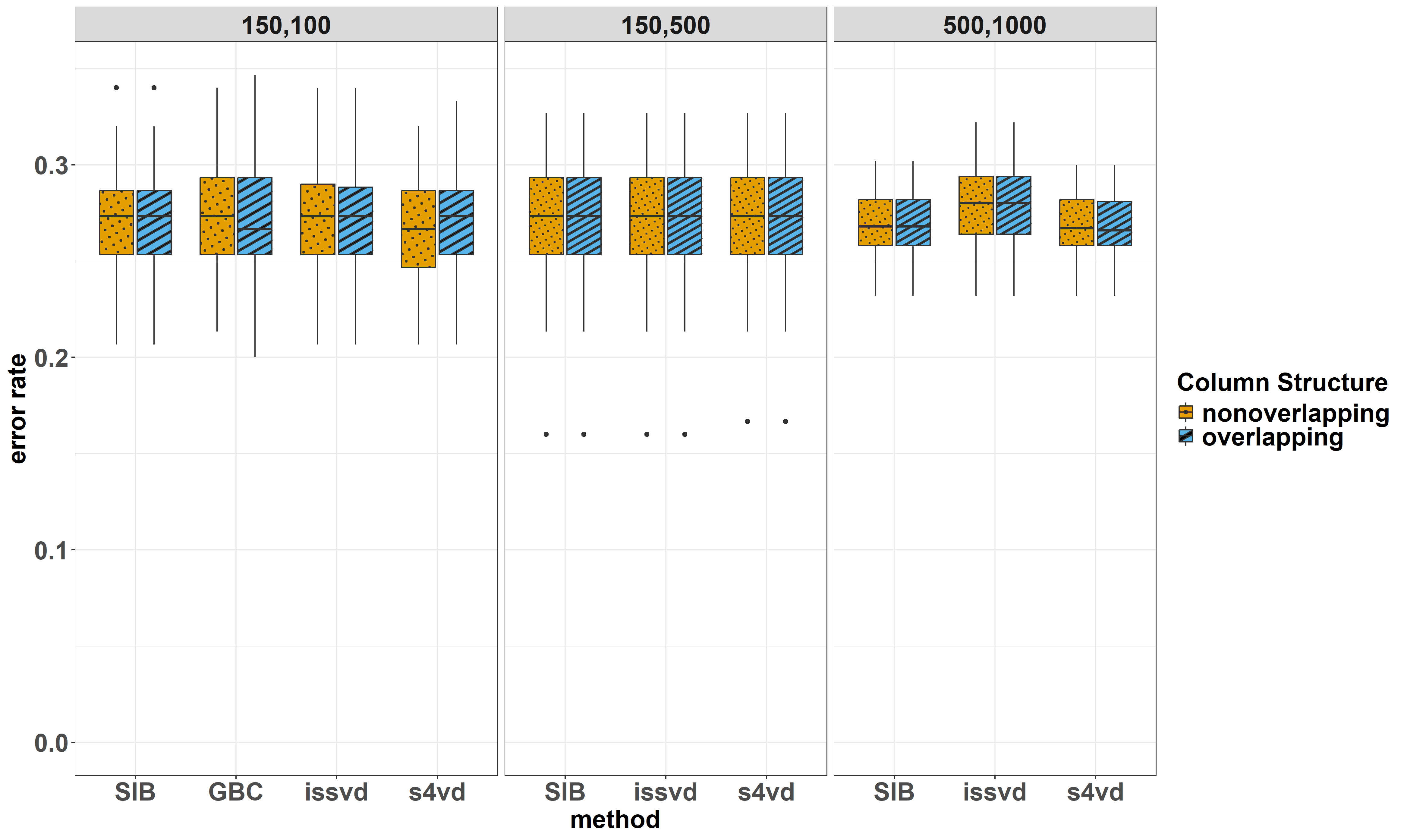}
\caption{Boxplot of the training error rate for all methods. The title indicates the number of samples and variables per view. The color of the boxes indicates whether or not we assume the columns to be nonoverlapping when estimating the biclusters.}
\end{figure}

\begin{figure}[H]
\includegraphics[width=\textwidth]{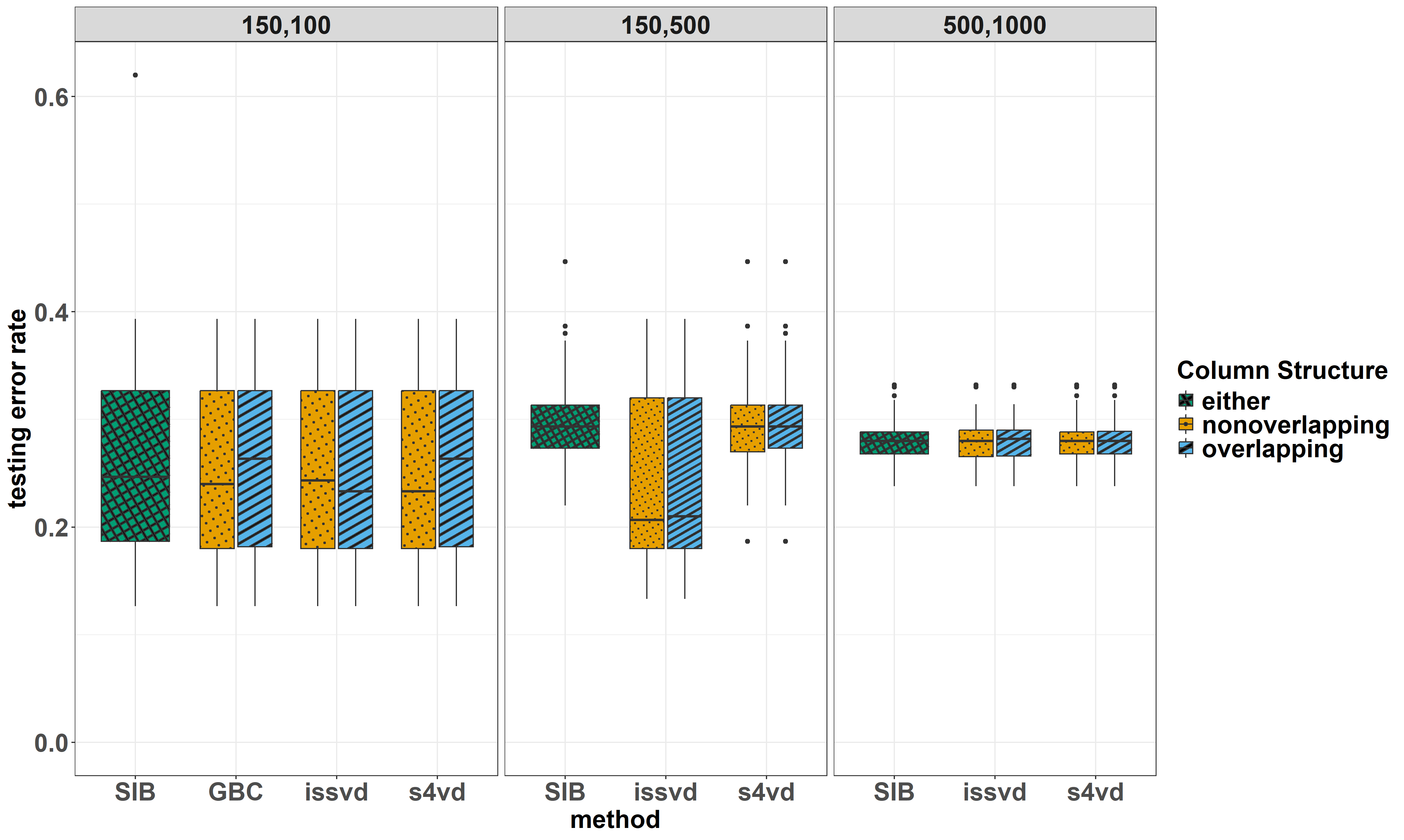}
\caption{Boxplot of the testing error rate for all methods. The title indicates the number of samples and variables per view. The color of the boxes indicates whether or not we assume the columns to be nonoverlapping when estimating the biclusters. Since the prediction with our proposed method does not depend on column structures, it is labeled as "either".}
\end{figure}

\begin{longtable}[c]{|c|c|c|c|c|}
\toprule
$n$ & $p$ & Methods & Time (minutes:seconds) \\
\midrule
150 & 100 & SIB & 2:20\\
150 & 100 & GBC & 37:57\\
150 & 100 & iSSVD & 0:08\\
150 & 100 & S4VD & 0:16\\
\midrule
150 & 500 & SIB & 3:22\\
150 & 500 & iSSVD & 0:07\\
150 & 500 & S4VD & 0:13\\
\midrule
500 & 1000  & SIB & 13:21\\
500 & 1000  & iSSVD & 0:13\\
500 & 1000  & S4VD & 0:48\\
\bottomrule
\caption{Computational times for all the methods in simulation settings. The time reported is the average computational time of the 100 repetitions for each method. To make the table more readable, we took the average of different outcome types for the proposed method, and we took the average of nonoverlapping and overlapping column structures for iSSVD and S4VD. The time for the proposed method with different outcomes is similar, and the time for iSSVD and S4VD under different assumptions of the column structure is similar. We did not include the running time for GBC with larger $n$ and $p$ because the running time is too long, making it impractical. All the methods are executed on a cloud server with the same number of cores (21 cores) to ensure the fairness of the comparison. The proposed method and iSSVD are written in Python, while GBC and S4VD are written in Python. }
\end{longtable}

\section{Additional Results in Real Data Analysis}

\par Web Table C.1 summarizes the association of all the clinical characteristics considered and the biclusters identified by the proposed method. We list all the lipids and imaging variables selected for each bicluster in Web Tables C.2 and C.3, respectively. We present the association of clinical variables and biclusters identified by iSSVD in the Web Table C.4.
\begin{longtable}{@{\extracolsep{\fill}}lcccc}
\toprule
\textbf{Bicluster} & \textbf{1}  & \textbf{2} & \textbf{3}& \\
& N = 84 &N = 334 & N = 62 & \textbf{p-value}\\ 
\midrule\addlinespace[2.5pt]
\textbf{Demographical Information} &  &  &  & \\
Age & 71 (67, 77) & 72 (68, 77) & 75 (68, 79) & 0.15 \\ 
Gender &  &  &  & 0.11 \\ 
    Female & 33 (39\%) & 174 (52\%) & 30 (48\%) &  \\ 
    Male & 51 (61\%) & 160 (48\%) & 32 (52\%) &  \\ 
Ethnicity &  &  &  & 0.3 \\ 
    Hisp/Latino & 0 (0\%) & 11 (3.3\%) & 3 (4.8\%) &  \\ 
    Not Hisp/Latino & 84 (100\%) & 322 (96\%) & 59 (95\%) &  \\ 
    Unknown & 0 (0\%) & 1 (0.3\%) & 0 (0\%) &  \\ 
Race &  &  &  & 0.4 \\ 
    Am Indian/Alaskan & 0 (0\%) & 1 (0.3\%) & 0 (0\%) &  \\ 
    Asian & 1 (1.2\%) & 6 (1.8\%) & 0 (0\%) &  \\ 
    Black & 4 (4.8\%) & 15 (4.5\%) & 4 (6.5\%) &  \\ 
    Hawaiian/Other PI & 1 (1.2\%) & 0 (0\%) & 0 (0\%) &  \\ 
    More than one & 1 (1.2\%) & 2 (0.6\%) & 0 (0\%) &  \\ 
    Unknown & 0 (0\%) & 0 (0\%) & 1 (1.6\%) &  \\ 
    White & 77 (92\%) & 310 (93\%) & 57 (92\%) &  \\ 
Education & 16.00 (14.00, 18.00) & 17.00 (14.00, 18.00) & 16.00 (14.00, 18.00) & 0.044 \\ 
Type of Participant residence &  &  &  & 0.7 \\ 
    Apartment (rented) & 4 (4.8\%) & 19 (5.7\%) & 5 (8.1\%) &  \\ 
    Assisted Living & 0 (0\%) & 0 (0\%) & 1 (1.6\%) &  \\ 
    Condo/Co-op (owned) & 13 (15\%) & 41 (12\%) & 7 (11\%) &  \\ 
    House & 63 (75\%) & 255 (76\%) & 46 (74\%) &  \\ 
    Mobile Home & 1 (1.2\%) & 1 (0.3\%) & 0 (0\%) &  \\ 
    Other (specify) & 0 (0\%) & 6 (1.8\%) & 1 (1.6\%) &  \\ 
    Retirement Community & 3 (3.6\%) & 12 (3.6\%) & 2 (3.2\%) &  \\ 
Normal activities of daily living &  &  &  & 0.7 \\ 
    Marginal & 5 (11\%) & 15 (8.8\%) & 3 (17\%) &  \\ 
    No & 0 (0\%) & 3 (1.8\%) & 0 (0\%) &  \\ 
    Yes & 41 (89\%) & 152 (89\%) & 15 (83\%) &  \\ 
    Unknown & 38 & 164 & 44 &  \\ 
Dropped activities and interests &  &  &  & 0.044 \\ 
    No(0) & 70 (83\%) & 305 (91\%) & 52 (84\%) &  \\ 
    Yes(1) & 14 (17\%) & 29 (8.7\%) & 10 (16\%) &  \\ 
Currently smoking & 40 (48\%) & 126 (38\%) & 21 (34\%) & 0.2 \\ 
Alcohol abuse & 3 (3.6\%) & 10 (3.0\%) & 7 (11\%) & 0.016 \\ 
\midrule
\textbf{Clinical Comorbidities} &&&&\\
Diabetes & 4 (13\%) & 14 (8.8\%) & 1 (10\%) & 0.6 \\ 
    Unknown & 54 & 174 & 52 &  \\ 
Hypertension & 15 (50\%) & 58 (36\%) & 4 (40\%) & 0.4 \\ 
    Unknown & 54 & 174 & 52 &  \\ 
Dyslipidemia & 4 (13\%) & 39 (24\%) & 1 (10\%) & 0.3 \\ 
    Unknown & 54 & 174 & 52 &  \\ 
Emotional Incontinence &  &  &  & 0.053 \\ 
    Absent & 83 (99\%) & 333 (100\%) & 60 (97\%) &  \\ 
    Present - 1 point & 1 (1.2\%) & 1 (0.3\%) & 2 (3.2\%) &  \\ 
\midrule
\textbf{AD Related Conditions} &&&&\\
Family History of AD & 26 (31\%) & 126 (38\%) & 19 (31\%) & 0.3 \\ 
Diagnosis (3-level) &  &  &  & <0.001 \\ 
    CN & 22 (26\%) & 150 (45\%) & 5 (8.1\%) &  \\ 
    MCI & 47 (56\%) & 170 (51\%) & 18 (29\%) &  \\
    AD & 15 (18\%) & 14 (4.2\%) & 39 (63\%) &  \\ 
Diagnosis (5-level) &  &  &  & <0.001 \\ 
    CN & 16 (19\%) & 115 (34\%) & 4 (6.5\%) &  \\ 
    SMC & 6 (7.1\%) & 35 (10\%) & 1 (1.6\%) &  \\ 
    EMCI & 22 (26\%) & 91 (27\%) & 7 (11\%) &  \\ 
    LMCI & 25 (30\%) & 79 (24\%) & 11 (18\%) &  \\ 
    AD & 15 (18\%) & 14 (4.2\%) & 39 (63\%) &  \\ 
MMSE & 27 (26, 29) & 29 (28, 30) & 23 (22, 25) & <0.001 \\ 
CDRSB & 1.50 (0.00, 2.50) & 0.50 (0.00, 1.50) & 3.50 (2.00, 5.00) & <0.001 \\ 
ADAS13 & 17 (11, 22) & 11 (7, 16) & 27 (20, 33) & <0.001 \\ 
    Unknown & 0 & 1 & 1 &  \\ 
mPACCdigit & -7 (-11, -3) & -2 (-5, 0) & -16 (-19, -12) & <0.001 \\ 
mPACCtrailsB & -6.2 (-9.1, -1.6) & -1.7 (-4.6, 0.6) & -14.3 (-16.1, -11.1) & <0.001 \\ 
\midrule
\textbf{AD Biomarkers} &&&&\\
Hippocampus & 6,970 (5,903, 8,054) & 7,291 (6,679, 7,864) & 6,030 (5,111, 6,702) & <0.001 \\ 
    Unknown & 8 & 31 & 5 &  \\ 
APOE4 &  &  &  & <0.001 \\ 
    0 & 35 (42\%) & 201 (60\%) & 22 (35\%) &  \\ 
    1 & 36 (43\%) & 108 (32\%) & 28 (45\%) &  \\ 
    2 & 13 (15\%) & 25 (7.5\%) & 12 (19\%) &  \\ 
\bottomrule
\caption{Baseline characteristics of ADNI participants by the biclusters identified with the proposed method. Values are median (first and third quartiles) for continuous variables and N (percentages) for binary/categorical variables. P-values are calculated by comparing all three biclusters using the Kruskal-Wallis test for continuous variables, and the Chi-square test or Fisher's exact test for binary/categorical variables, depending on the sizes of the cells.}
\end{longtable}

\begin{longtable}{l|l|l}
\toprule
Bicluster 1 & Bicluster 2 & Bicluster 3\\
\midrule
CER.D16.1.20.0. & CER1P.D18.1.16.0. & CER.D19.1.24.0.\\
CER.D16.1.23.0. & HEXCER.D18.2.22.0. & PC.16.0\_20.5.\\
CER.D16.1.24.0. & HEXCER.D18.2.24.0. & PC.16.1\_22.6.\\
CER.D16.1.24.1. & HEX2CER.D18.1.22.0. & PC.18.0\_22.6.\\
CER.D17.1.23.0. & HEX2CER.D18.1.24.1. & PC.36.6...A.\\
CER.D17.1.24.1. & HEX2CER.D18.2.24.1. & PC.38.5...B.\\
CER.D18.1.14.0. & HEX3CER.D18.1.16.0. & PC.O.16.0.20.3.\\
CER.D18.1.16.0. & HEX3CER.D18.1.18.0. & PC.O.18.0.20.4.\\
CER.D18.1.23.0. & HEX3CER.D18.1.20.0. & PC.O.35.4.\\
CER.D18.2.14.0. & HEX3CER.D18.1.22.0. & PC.O.36.0.\\
CER.D18.2.23.0. & HEX3CER.D18.1.24.0. & PC.P.15.0.20.4...A.\\
CER.M18.0.23.0. & HEX3CER.D18.1.24.1. & PC.P.15.0.20.4...B.\\
SM.D17.1.14.0. & GM3.D18.1.16.0. & PC.P.17.0.20.4...A.\\
SM.D17.1.16.0. & GM3.D18.1.24.0. & PC.P.17.0.20.4...B.\\
SM.D18.0.14.0. & SULFATIDE..D18.1..24.1. & PC.P.35.2..A.\\
SM.D18.1.14.0..SM.D16.1.16.0. & SM.35.2...B. & PC.P.36.3.\\
SM.D18.1.18.0..SM.D16.1.20.0. & SM.44.3. & LPC.18.2...SN1.\\
SM.D18.2.14.0. & SM.D18.0.16.0. & LPC.18.2...SN2.\\
PC.14.0\_16.0. & SM.D18.2.24.0. & LPC.O.18.0.\\
PC.14.0\_20.4. & PC.O.18.0.18.1. & LPC.P.17.0...A.\\
PC.14.0\_22.6. & PC.O.18.0.18.2. & PE.16.0\_22.6.\\
PC.15.MHDA\_18.1. & PC.O.18.1.18.1. & PE.18.0\_22.6.\\
PC.15.MHDA\_18.2. & PC.O.18.1.18.2. & PE.18.1\_22.6...A.\\
PC.15.MHDA\_20.4. & PC.O.34.1. & PE.O.16.0.22.4.\\
PC.15.0\_20.3. & PC.O.34.2. & PE.P.15.0.20.4...A.\\
PC.15.0\_20.4. & PC.P.16.0.18.0. & PE.P.15.0.20.4...B.\\
PC.16.0\_16.0. & PC.P.16.0.18.1. & PE.P.16.0.18.1.\\
PC.16.0\_18.0. & PC.P.16.0.18.2. & PE.P.16.0.18.2.\\
PC.16.0\_18.3...A. & PC.P.16.0.18.3. & PE.P.16.0.20.3...A.\\
PC.16.0\_18.3...B. & PC.P.18.0.18.2. & PE.P.16.0.20.3...B.\\
PC.16.0\_20.3...A. & PC.P.18.0.22.5. & PE.P.16.0.22.4.\\
PC.16.0\_20.3...B. & PC.P.18.1.18.1. & PE.P.18.0.18.1.\\
PC.16.1\_20.4. & PC.P.18.1.22.6. & PE.P.18.0.18.2.\\
PC.17.0\_18.1. & PC.P.35.2..B. & PE.P.18.0.20.3...A.\\
PC.18.0\_18.1. & PC.P.38.5...A. & PE.P.18.0.20.3...B.\\
PC.18.0\_20.3. & PG.36.1. & PE.P.18.0.22.4.\\
PC.18.1\_20.3. & PG.36.2. & PE.P.18.0.22.5...N6.\\
PC.28.0. & CE.16.2. & PE.P.18.1.18.2...B.\\
PC.31.0...A. & CE.18.3. & PE.P.18.1.20.3...A.\\
PC.31.0...B. & CE.20.4. & PE.P.18.1.20.3...B.\\
PC.32.1. & CE.22.6. & FA.20.5.\\
PC.32.2. & DE.18.2. & AC.26.0.\\
PC.33.0...A. & DE.20.4. & DG.18.1\_20.5.\\
PC.33.0...B. & METHYL.CE.22.6. & DG.18.2\_22.6.\\
PC.33.1. & METHYL.DE.18.2. & TG.54.7...SIM.\\
PC.38.2. & DG.14.0\_16.0. & TG.56.7...SIM.\\
PC.38.4...B. & DG.14.0\_18.2. & TG.56.8...SIM.\\
PC.38.6...A. & DG.16.0\_16.0. & TG.56.9...SIM.\\
PC.39.5..B. & DG.16.0\_16.1. & TG.58.10...SIM.\\
PC.40.7...A. & DG.16.0\_18.1. & TG.58.8...SIM.\\
PC.40.8. & DG.16.0\_18.2. & TG.58.9...SIM.\\
LPC.14.0...SN1. & DG.16.0\_20.4. & TG.54.7...NL.20.5.\\
LPC.14.0...SN2. & DG.16.1\_18.1. & TG.56.7...NL.20.5.\\
LPC.15.MHDA...SN1...104\_SN1. & DG.18.0\_18.1. & TG.56.7...NL.22.5.\\
LPC.15.MHDA...SN1....LPC.17.0...SN2. & DG.18.0\_18.2. & TG.56.8...NL.20.5.\\
LPC.15.MHDA...SN2. & DG.18.1\_18.1. & TG.56.8...NL.22.6.\\
LPC.15.0...SN1. & DG.18.1\_20.4. & TG.56.9...NL.22.6.\\
LPC.15.0...SN2. & TG.48.1...SIM. & TG.58.10...NL.22.6.\\
LPC.16.0...SN1. & TG.48.3...SIM. & TG.58.8...NL.22.6.\\
LPC.16.0...SN2. & TG.50.0...SIM. & TG.58.9...NL.22.6.\\
LPC.16.1...SN1. & TG.50.1...SIM. & TG.O.50.1...NL.15.0.\\
LPC.16.1...SN2. & TG.50.2...SIM. & TG.O.50.1...NL.16.0.\\
LPC.17.1...A...SN1...104\_SN1. & TG.50.3...SIM. & TG.O.50.1...NL.18.1.\\
LPC.17.1...SN1...A....LPC.17.1...SN2...B. & TG.50.4...SIM. & TG.O.52.1...NL.16.0.\\
LPC.17.1...SN1...B. & TG.51.1...SIM. & TG.O.52.1...NL.18.1.\\
LPC.17.1...SN2...A. & TG.51.2...SIM. & TG.O.52.2...NL.17.1.\\
LPC.20.3...SN2. & TG.52.1...SIM. & TG.O.52.2...NL.18.1.\\
PE.15.MHDA\_18.1. & TG.52.2...SIM. & TG.O.54.2...NL.17.1.\\
PE.15.MHDA\_18.2. & TG.53.2...SIM. & TG.O.54.2...NL.18.1.\\
PE.15.MHDA\_20.4. & TG.54.1...SIM. &\\
PE.15.MHDA\_22.6. & TG.54.2...SIM. &\\
PE.16.0\_16.0. & TG.48.1...NL.18.1. &\\
PE.16.0\_16.1. & TG.48.2...NL.14.0. &\\
PE.16.0\_18.3...A. & TG.48.2...NL.18.2. &\\
PE.16.0\_18.3...B. & TG.48.3...NL.18.3. &\\
PE.16.0\_20.4. & TG.50.0...NL.18.0. &\\
PE.16.1\_18.2. & TG.50.1...NL.16.0. &\\
PE.16.1\_20.4. & TG.50.1...NL.18.1. &\\
PE.18.0\_20.3...B. & TG.50.2...NL.18.1. &\\
PE.18.0\_22.5...N3. & TG.50.3...NL.14.0. &\\
PE.18.0\_22.5...N6. & TG.50.3...NL.16.1. &\\
PE.38.5...B. & TG.50.3...NL.18.2. &\\
LPE.16.0...SN1. & TG.50.3...NL.18.3. &\\
LPE.16.0...SN2. & TG.50.4...NL.18.3. &\\
LPE.18.0...SN1. & TG.51.1...NL.17.0. &\\
LPE.18.0...SN2. & TG.51.2...NL.17.0. &\\
PI..38.5...B. & TG.51.2...NL.17.1. &\\
PI.15.MHDA\_20.4..PI.17.0\_20.4. & TG.52.1...NL.18.0. &\\
PI.16.0\_16.1. & TG.52.1...NL.18.1. &\\
PI.16.0\_20.3...B. & TG.52.2...NL.16.0. &\\
PI.16.0\_20.4. & TG.52.2...NL.18.2. &\\
PI.16.0.16.0. & TG.52.4...NL.18.3. &\\
PI.18.0\_20.4. & TG.53.2...NL.17.1. &\\
PI.18.0\_22.4. & TG.53.2...NL.18.1. &\\
PI.34.1. & TG.54.1...NL.18.1. &\\
PI.38.5...A. & TG.54.2...NL.18.0. &\\
PG.34.1. & TG.54.2...NL.20.1. &\\
CE.17.0. & TG.54.4...NL.20.3. &\\
CE.17.1. & TG.54.5...NL.20.4. &\\
CE.18.0. & &\\
TG.48.2...SIM. & &\\
TG.48.2...NL.14.1. & &\\
TG.48.2...NL.16.1. & &\\
TG.48.3...NL.14.0. & &\\
TG.48.3...NL.16.1. & &\\
TG.49.1...NL.17.1. & &\\
TG.50.2...NL.16.1. & &\\
TG.50.4...NL.20.4. & &\\
TG.52.5...NL.20.4. & &\\
TG.O.52.0...NL.16.0. & &\\
PC.34.2....OH. & &\\
\bottomrule
\caption{Lipidomics variables identified in each bicluster. The abbreviation is used for simplicity.}
\end{longtable} 

\setlength\LTleft{-0.5in}
\begin{longtable}{|p{6cm}|p{6cm}|p{6cm}|}
\toprule
Bicluster 1 & Bicluster 2 & Bicluster 3\\
\midrule
\raggedright Intracranial CSF volume & \raggedright Surface Area of RightParahippocampal &  Volume (Cortical Parcellation) of RightPrecuneus\\ \hline
\raggedright Volume (Cortical Parcellation) of Icv & \raggedright Volume (Cortical Parcellation) of RightParsOrbitalis &  Volume (Cortical Parcellation) of RightInferiorParietal\\ \hline 
\raggedright Surface Area of RightPrecuneus & Surface Area of RightParsOrbitalis & Volume (Cortical Parcellation) of RightInferiorTemporal\\ 
\hline 
\raggedright Volume (Cortical Parcellation) of RightRostralMiddleFrontal &\raggedright Volume (Cortical Parcellation) of RightPericalcarine & Subcortical Volume (aseg.stats) of CorticalGM\\ 
\hline 
\raggedright Surface Area of RightRostralMiddleFrontal &\raggedright Surface Area of RightPericalcarine & Surface Area (aparc.stats) of LeftHemisphereWM\\ 
\hline 
\raggedright Volume (Cortical Parcellation) of RightSuperiorFrontal &\raggedright Surface Area of RightPosteriorCingulate & Surface Area (aparc.stats) of RightHemisphereWM\\ 
\hline 
\raggedright Surface Area of RightSuperiorParietal &\raggedright Surface Area of RightPrecentral & Subcortical Volume (aseg.stats) of LeftCorticalGM\\ 
\hline 
\raggedright Volume (WM Parcellation) of RightVentralDC &\raggedright Surface Area of RightSuperiorFrontal & Subcortical Volume (aseg.stats) of RightCorticalGM\\ 
\hline 
\raggedright Volume (WM Parcellation) of Brainstem &\raggedright Surface Area of LeftFusiform & Subcortical Volume (aseg.stats) of LeftCorticalWM\\ 
\hline 
\raggedright Volume (Cortical Parcellation) of LeftIsthmusCingulate &\raggedright Volume (Cortical Parcellation) of LeftLateralOrbitofrontal & Subcortical Volume (aseg.stats) of CorticalWM\\ 
\hline 
\raggedright Surface Area of LeftIsthmusCingulate &\raggedright Surface Area of LeftLateralOrbitofrontal & Subcortical Volume (aseg.stats) of TotalGM\\ 
\hline 
\raggedright Surface Area of LeftLateralOccipital &\raggedright Surface Area of LeftParsOrbitalis & Subcortical Volume (aseg.stats) of SupraTentorial\\ 
\hline 
\raggedright Volume (Cortical Parcellation) of LeftMedialOrbitofrontal &\raggedright Volume (Cortical Parcellation) of LeftSuperiorFrontal &\\ 
\hline 
\raggedright Surface Area of LeftMedialOrbitofrontal &\raggedright Surface Area of LeftSuperiorFrontal &\\ 
\hline 
\raggedright Surface Area of LeftMiddleTemporal &\raggedright Surface Area of LeftSuperiorTemporal &\\ 
\hline 
\raggedright Surface Area of LeftPostcentral &\raggedright Surface Area of RightFusiform &\\ 
\hline 
\raggedright Surface Area of LeftPrecuneus &\raggedright Surface Area of RightInferiorParietal &\\ 
\hline 
\raggedright Volume (Cortical Parcellation) of LeftRostralMiddleFrontal &\raggedright Surface Area of RightInferiorTemporal &\\ 
\hline 
\raggedright Surface Area of LeftRostralMiddleFrontal &\raggedright Surface Area of RightLateralOrbitofrontal &\\ 
\hline 
\raggedright Surface Area of LeftSupramarginal &\raggedright Volume (Cortical Parcellation) of RightMedialOrbitofrontal &\\ 
\hline 
\raggedright Volume (WM Parcellation) of LeftVentralDC &\raggedright Surface Area of RightMedialOrbitofrontal &\\ 
\hline 
\raggedright Surface Area of RightCaudalMiddleFrontal &\raggedright Surface Area of RightMiddleTemporal &\\ 
\hline 
\raggedright Surface Area of RightIsthmusCingulate &\raggedright Volume (Cortical Parcellation) of RightInsula &\\ 
\hline 
\raggedright Volume (Cortical Parcellation) of RightLateralOccipital & &\\ 
\hline 
\raggedright Surface Area of RightLateralOccipital & &\\ 
\hline 
\raggedright Volume (Cortical Parcellation) of RightLateralOrbitofrontal & &\\ 
\hline 
\raggedright Surface Area of LeftInsula & &\\ 
\hline 
\raggedright Surface Area of RightInsula & &\\ 
\hline 
\raggedright Subcortical Volume (aseg.stats) of RightCorticalWM & &\\ 
\hline 
\raggedright Subcortical Volume (aseg.stats) of SubcorticalGM & &\\ 
\bottomrule

\caption{Image numeric variables identified in each bicluster.}
\end{longtable}

\setlength\LTleft{0in}
\setlength\LTright{1in}
\begin{longtable}[t]{@{\extracolsep{\fill}}lcccc}
\toprule
\textbf{Bicluster} & \textbf{1}  & \textbf{2} & \textbf{3}& \\
& N = 200 &N = 224 & N = 56 & \textbf{p-value}\\ 
\midrule\addlinespace[2.5pt]
\textbf{Demographical Information} &  &  &  & \\
Age & 73 (68, 77) & 71 (67, 76) & 74 (68, 78) & 0.2 \\ 
Gender &  &  &  & 0.059 \\ 
    Female & 94 (47\%) & 107 (48\%) & 36 (64\%) &  \\ 
    Male & 106 (53\%) & 117 (52\%) & 20 (36\%) &  \\ 
Ethnicity &  &  &  & 0.5 \\ 
    Hisp/Latino & 5 (2.5\%) & 6 (2.7\%) & 3 (5.4\%) &  \\ 
    Not Hisp/Latino & 194 (97\%) & 218 (97\%) & 53 (95\%) &  \\ 
    Unknown & 1 (0.5\%) & 0 (0\%) & 0 (0\%) &  \\ 
Race &  &  &  & 0.7 \\ 
    Am Indian/Alaskan & 1 (0.5\%) & 0 (0\%) & 0 (0\%) &  \\ 
    Asian & 5 (2.5\%) & 1 (0.4\%) & 1 (1.8\%) &  \\ 
    Black & 9 (4.5\%) & 10 (4.5\%) & 4 (7.1\%) &  \\ 
    Hawaiian/Other PI & 0 (0\%) & 1 (0.4\%) & 0 (0\%) &  \\ 
    More than one & 1 (0.5\%) & 2 (0.9\%) & 0 (0\%) &  \\ 
    Unknown & 0 (0\%) & 1 (0.4\%) & 0 (0\%) &  \\ 
    White & 184 (92\%) & 209 (93\%) & 51 (91\%) &  \\ 
Education & 17.00 (14.00, 18.00) & 16.00 (14.00, 18.00) & 16.50 (14.50, 18.50) & 0.14 \\ 
Type of Participant residence &  &  & & $0.34^1$\\ 
    Apartment (rented) & 13 (6.5\%) & 9 (4.0\%) & 6 (11\%) &\\ 
    Assisted Living & 0 (0\%) & 0 (0\%) & 1 (1.8\%) &\\ 
    Condo/Co-op (owned) & 26 (13\%) & 28 (13\%) & 7 (13\%) &\\ 
    House & 152 (76\%) & 174 (78\%) & 38 (68\%) &\\ 
    Mobile Home & 1 (0.5\%) & 1 (0.4\%) & 0 (0\%) &\\ 
    Other (specify) & 1 (0.5\%) & 4 (1.8\%) & 2 (3.6\%) &\\ 
    Retirement Community & 7 (3.5\%) & 8 (3.6\%) & 2 (3.6\%) &\\ 
Normal activities of daily living &  &  &  & >0.9 \\ 
    Marginal & 10 (10\%) & 11 (9.8\%) & 2 (7.7\%) &  \\ 
    No & 1 (1.0\%) & 2 (1.8\%) & 0 (0\%) &  \\ 
    Yes & 85 (89\%) & 99 (88\%) & 24 (92\%) &  \\ 
    Unknown & 104 & 112 & 30 &  \\ 
Dropped activities and interests &  &  &  & 0.5 \\ 
    No(0) & 175 (88\%) & 200 (89\%) & 52 (93\%) &  \\ 
    Yes(1) & 25 (13\%) & 24 (11\%) & 4 (7.1\%) &  \\ 
Currently smoking & 82 (41\%) & 84 (38\%) & 21 (38\%) & 0.7 \\ 
Alcohol abbuse & 8 (4.0\%) & 9 (4.0\%) & 3 (5.4\%) & 0.9 \\ 
\midrule
\textbf{Clinical Comorbidities} &&&&\\
Diabetes & 11 (14\%) & 7 (7.6\%) & 1 (3.7\%) & 0.3 \\ 
    Unknown & 119 & 132 & 29 &  \\ 
Hypertension & 34 (42\%) & 37 (40\%) & 6 (22\%) & 0.2 \\ 
    Unknown & 119 & 132 & 29 &  \\ 
Dyslipidemia & 16 (20\%) & 22 (24\%) & 6 (22\%) & 0.8 \\ 
    Unknown & 119 & 132 & 29 &  \\ 
Emotional Incontinence &  &  &  & >0.9 \\ 
    Absent & 198 (99\%) & 222 (99\%) & 56 (100\%) &  \\ 
    Present - 1 point & 2 (1.0\%) & 2 (0.9\%) & 0 (0\%) &  \\ 
\midrule
\textbf{AD Related Conditions} &&&&\\
Family History of AD & 74 (37\%) & 72 (32\%) & 25 (45\%) & 0.2 \\ 
Diagnosis (3-level) &  &  &  & 0.8 \\ 
    CN & 74 (37\%) & 79 (35\%) & 24 (43\%) &  \\ 
    MCI & 96 (48\%) & 113 (50\%) & 26 (46\%) &  \\ 
    AD & 30 (15\%) & 32 (14\%) & 6 (11\%) &  \\ 
Diagnosis (5-level) &  &  &  & $0.78^1$\\ 
    CN & 53 (27\%) & 63 (28\%) & 19 (34\%) &\\
    SMC & 21 (11\%) & 16 (7.1\%) & 5 (8.9\%) &\\ 
    EMCI & 49 (25\%) & 61 (27\%) & 10 (18\%) &\\ 
    LMCI & 47 (24\%) & 52 (23\%) & 16 (29\%) &\\ 
    AD & 30 (15\%) & 32 (14\%) & 6 (11\%) &\\ 
MMSE & 29 (26, 30) & 29 (26, 30) & 29 (27, 30) & 0.5 \\ 
CDRSB & 1.00 (0.00, 2.50) & 1.00 (0.00, 2.50) & 0.50 (0.00, 2.00) & 0.4 \\ 
ADAS13 & 12 (9, 20) & 14 (9, 20) & 10 (6, 20) & 0.2 \\ 
    Unknown & 1 & 0 & 1 &  \\ 
mPACCdigit & -3 (-10, 0) & -4 (-9, -1) & -2 (-9, 1) & 0.2 \\ 
mPACCtrailsB & -2.8 (-8.1, -0.4) & -3.6 (-7.8, -0.4) & -1.5 (-7.0, 1.3) & 0.2 \\ 
\midrule
\textbf{AD Biomarkers} &&&&\\
Hippocampus & 7,086 (6,053, 7,810) & 7,197 (6,569, 7,848) & 6,987 (6,259, 7,851) & 0.5 \\ 
    Unknown & 20 & 21 & 3 &  \\ 
APOE4 &  &  &  & 0.7 \\ 
    0 & 105 (53\%) & 125 (56\%) & 28 (50\%) &  \\ 
    1 & 72 (36\%) & 76 (34\%) & 24 (43\%) &  \\ 
    2 & 23 (12\%) & 23 (10\%) & 4 (7.1\%) &  \\ 
\bottomrule
\caption{Baseline characteristics of ADNI participants by the biclusters identified with iSSVD. Values are median (first and thrid quartiles) for continuous variables, and N (percentages) for binary/categorical variables. P-values are calculated by comparing all three biclusters using Kruskal-Wallis test for continuous variables, and Chi-square test or Fisher's exact test for binary/categorical variables depending on the sizes of the cells.\\ 
\textsuperscript{\textit{1}} The p-values for type of residence and five-level diagnosis are simulated due to the size of the cells.}
\end{longtable}

\section{Algorithms}
\begin{algorithm}[H]
	\caption{Estimate all the parameters} 
    \hspace*{\algorithmicindent} \textbf{Input:} $\bm{X}$, $\bm{y}$, $t$ (tolerance), $\alpha$, $M$ (maximum number of iterations) \\
    \hspace*{\algorithmicindent} \textbf{Output: $\bm{U}$, $\bm{V}$, $\bm{W}$, $\bm{\mu}$, $\bm{\beta}$} 
	\begin{algorithmic}[1]
            \State Initialize $\bm{U}$, $\bm{V}$, $\bm{W}$, $\bm{\mu}$, and $\bm{\beta}$
            \State Calculate the value of the differentiable part of the loss function $l(\bm{X,y,U,V,W,\mu,\beta})$
        \For{$m \gets 1$ to $M$}
                \State Set $\bm{U}_{unproj} = (\bm{U} - \alpha * \nabla l/\nabla \bm{U})$, and update columns of $\bm{U} = \bm{U}_{unproj}/||\bm{U}_{unproj}||^2$
                \State Update $l_{new}(\bm{X,y,U,V,W,\mu,\beta})$
                \State Set $\bm{W}_{unproj} = (\bm{W} - \alpha * \nabla l_{new}/\nabla \bm{W})$, and update $\bm{W}$ by projecting rows of $\bm{W}_{unproj}$ onto a probability simplex
                \State Update $l_{new}(\bm{X,y,U,V,W,\mu,\beta})$
                 \State Set $\bm{V}_{unproj} = (\bm{V} - \alpha * \nabla l_{new}/\nabla \bm{V})$, and update $\bm{V}$ by applying the proximal map for Lasso on each entry of $\bm{V}_{unproj}$
                 \State Update $l_{new}(\bm{X,y,U,V,W,\mu,\beta})$
                  \State Update $\bm{\mu} = (\bm{\mu} - \alpha * \nabla l_{new}/\nabla \bm{\mu})$
                  \State Update $l_{new}(\bm{X,y,U,V,W,\mu,\beta})$
                  \State Update $\bm{\beta} = (\bm{\beta} - \alpha * \nabla l_{new}/\nabla \bm{\beta})$
                  \State Update $l_{new}(\bm{X,y,U,V,W,\mu,\beta})$
                  \If{$\frac{|l_{new}(X,y,U,V,W,\mu,\beta) - l(X,y,U,V,W,\mu,\beta)|}{l(X,y,U,V,W,\mu,\beta)} < \text{t}$}
                    \State \textbf{break}
                  \EndIf
                  \If{any column of $V$ is all zero}
                    \State \textbf{break}
                  \EndIf
            \State Update $l(\bm{X, U,V,W,\mu}) = l_{new}(\bm{X, U,V,W,\mu})$
            \EndFor
            \State Update $\bm{\beta} = \text{argmin } \sum_{i=1}^n(\bm{y_i\Psi_{yi}}-G(\bm{\Psi_{yi}}))$
            \State \Return $\bm{U}$, $\bm{V}$, $\bm{W}$, $\bm{\mu}$, and $\bm{\beta}$
	\end{algorithmic} 
\end{algorithm}



\begin{algorithm}[H]
	\caption{Prediction}
    \hspace*{\algorithmicindent} \textbf{Input: } $X$, $V$, $\mu$, $t$ (tolerance), $\alpha$, $M$ (maximum number of iterations) \\
    \hspace*{\algorithmicindent} \textbf{Output: } $\bm{U}$, $\bm{W}$
	\begin{algorithmic}[1]
		  \State Initialize $\bm{U}$, and $\bm{W}$
            \State Calculate the value of the loss function $l_x(\bm{X, U,V,W,\mu})$
        \For{$m \gets 1$ to $M$}
                \State Set $\bm{U}_{unproj} = (\bm{U} - \alpha * \nabla l_x/\nabla \bm{U})$, and update columns of $\bm{U} = \bm{U}_{unproj}/||\bm{U}_{unproj}||^2$
                \State Update $l_{x, new}(\bm{X, U,V,W,\mu})$
                \State Set $\bm{W}_{unproj} = (\bm{W} - \alpha * \nabla l_{x, new}/\nabla \bm{W})$, and update $\bm{W}$ by projecting rows of $\bm{W}_{unproj}$ onto a probability simplex
                \State Update $l_{x, new}(\bm{X, U,V,W,\mu})$
                  \If{$\frac{|l_{x, new}(\bm{X, U,V,W,\mu}) - l_x(\bm{X, U,V,W,\mu})|}{l_x(\bm{X, U,V,W,\mu})} < \text{t}$}
                    \State \textbf{break}
                  \EndIf
            \State Update $l_x(\bm{X, U,V,W,\mu}) = l_{x, new}(\bm{X, U,V,W,\mu})$
            \EndFor
            \State \Return $\bm{U}$ and $\bm{W}$
	\end{algorithmic} 
\end{algorithm}

\bibliographystyle{unsrtnat} 
\bibliography{Supplementary.bib}